\documentclass[aps,pra,twocolumn,superscriptaddress,10pt]{revtex4-2} 
           
\usepackage{graphicx}%
\usepackage{amsmath,amssymb,amsfonts}%

\usepackage{siunitx}
\DeclareSIUnit\molar{\mole\per\cubic\deci\metre}
\DeclareSIUnit\Molar{M}

\begin{document}
\title{From shallow to full wrapping: geometry and deformability dictate lipid vesicle internalization}


\author{Stijn van der Ham}
\email{S.v.d.H contributed equally to this work with A.B.\\}
\affiliation{Active Soft Matter and Bio-inspired Materials Lab, Faculty of Science and Technology, MESA+ Institute, University of Twente, 7500 AE Enschede, The Netherlands}

\author{Alexander Brown}
\email{S.v.d.H contributed equally to this work with A.B.\\}
\affiliation{Department of Physics, Durham University, Durham, DH1 3LE, UK}

\author{Halim Kusumaatmaja}
\email{To whom correspondence should be addressed. \\ E-mail: h.r.vutukuri@utwente.nl \\ and halim.kusumaatmaja@ed.ac.uk.}
\affiliation{Institute for Multiscale Thermofluids, School of Engineering, The University of Edinburgh, Edinburgh, EH9 3FB, UK}

\author{Hanumantha Rao Vutukuri}
\email{To whom correspondence should be addressed. \\ E-mail: h.r.vutukuri@utwente.nl \\ and halim.kusumaatmaja@ed.ac.uk.}
\affiliation{Active Soft Matter and Bio-inspired Materials Lab, Faculty of Science and Technology, MESA+ Institute, University of Twente, 7500 AE Enschede, The Netherlands}

\begin{abstract}
The deformability of vesicles critically influences their engulfment by lipid membranes, a process central to endocytosis, viral entry, drug delivery, and intercellular transport. While theoretical models have long predicted this influence, direct experimental validation has remained elusive. Here, we combine experiments with continuum simulations to quantify how vesicle deformability affects the engulfment of small giant unilamellar vesicles (GUVs) by larger GUVs under depletion-induced adhesion. Using 3D confocal reconstructions, we extract vesicle shape, curvature, wrapping fraction, and the bendo-capillary length, a characteristic length scale that balances membrane bending and adhesion forces. We find that when vesicle size exceeds this length scale, engulfment is primarily governed by geometry. In contrast, when vesicle size is comparable to this scale, deformability strongly affects the transition between shallow, deep, and fully wrapped states, leading to suppression of full engulfment of vesicles. These findings connect theoretical predictions with direct measurements and offer a unified framework for understanding vesicle-mediated uptake across both synthetic and biological systems, including viral entry, synthetic cell design, drug delivery, and nanoparticle internalization.
\end{abstract}

\maketitle

Vesicle engulfment by lipid membranes is a fundamental mechanism by which cells internalize material, enabling cargo exchange across cellular membranes and playing essential roles in intercellular communication, immune responses, and intracellular transport \cite{gyorgy2011membrane}. In biological systems, extracellular vesicles such as exosomes, microvesicles, and apoptotic bodies transport proteins, lipids, and nucleic acids between cells, while pathogens like viruses exploit similar mechanisms to gain entry into host cells \cite{gyorgy2011membrane, arandjelovic2015phagocytosis, cohen2016viruses, marsh2006virus}. Examples include the phagocytosis of apoptotic bodies \cite{arandjelovic2015phagocytosis} and the cellular invasion by SARS-CoV-2 virions \cite{klein2020sars, bayati2021sars}. In therapeutic contexts, engineered liposomes mimic these uptake pathways to deliver drugs to target cells \cite{allen2013liposomal, guimaraes2021design, bareford2007endocytic}. While some vesicles cross membranes via direct fusion, many follow wrapping-based pathways such as endocytosis or phagocytosis, in which the vesicle is enclosed by the target cell’s plasma membrane before internalization or release \cite{joshi2020endocytosis, conner2003regulated, arandjelovic2015phagocytosis}. Understanding the physical principles that govern these processes is essential for interpreting cellular behaviour and for the development of more effective vesicle-based technologies.

The mechanics of vesicle engulfment are governed by the interplay between elastic membrane and adhesion forces. A central question is how these forces act on the object being engulfed. Most experimental and theoretical studies to date have primarily focused on the engulfment of rigid particles, where wrapping dynamics depend on factors such as adhesion strength, particle size and shape, membrane stiffness, spontaneous curvature, and the geometry of the engulfing membrane \cite{lipowsky1998vesicles, dasgupta2014shape, van2024role, bahrami2013orientational, agudo2015critical, agudo2021particle, spanke2020wrapping, bahrami2014wrapping}. In these systems, membrane deformation occurs around a fixed particle shape.

In contrast, vesicles are deformable, allowing the membrane and the vesicle to reshape each other during engulfment. Recent theoretical studies on the engulfment of soft objects, such as condensate droplets, soft particles and small lipid vesicles, by membranes have predicted that deformability can stabilize partially wrapped states \cite{yi2011cellular}, inhibit full engulfment \cite{yi2016incorporation}, and drive mutual shape remodelling between the object and the membrane \cite{imoto2022dynamin}. Moreover, these soft objects can be distinguished by their deformation modes, which can play an important role. For instance, in condensate droplets, soft particles, and lipid vesicles, these modes are respectively determined by the surface tension of the droplet, the bulk elasticity of the particle, and the bending deformation of the vesicle \cite{tang2016wrapping, midya2023membrane, satarifard2023mutual, Kusumaatmaja2011}. In parallel, experimental studies on nanoparticles demonstrated that particle elasticity and rigidity affect both cellular uptake and blood circulation time, emphasizing the broader biological relevance of deformability \cite{anselmo2015elasticity,sun2014tunable}. However, despite these advances, a systematic experimental investigation into the role of deformability in lipid vesicle engulfment remains elusive.

To address this gap, we conducted a combined experimental and theoretical study on the engulfment of small vesicles by larger ones, using giant unilamellar vesicles (GUVs) of varying sizes and low membrane tension as a model system \cite{dimova2019giant}. Our study demonstrates consistent morphologies between experiments and simulations for both endocytic and exocytic engulfment across a range of GUV size ratios, illustrating the diverse equilibrium states that emerge. The adhesion-driven interaction between these GUVs, counteracted by the membrane’s bending rigidity, was quantified using the bendo-capillary length. We find that when the small vesicle is much larger than this characteristic length scale, engulfment is primarily governed by geometric constraints, with deformability influencing only the minimum-energy shape. In contrast, when the small vesicle’s size is comparable to this scale, deformability plays a decisive role in determining the transition between partially and fully wrapped states.

\section*{Results}
\subsection*{Experimental realization of vesicle-vesicle engulfment}
To experimentally investigate vesicle engulfment, we use a biomimetic model system in which large GUVs engulf smaller ones, mimicking vesicle uptake by cells (see Supporting Movie S1 for an example of interacting vesicles). GUVs of varying sizes were prepared using the droplet transfer method \cite{dimova2019giant} (see Materials and Methods). Their membranes comprised DOPC lipids doped with \qty{0.1}{\mol\percent} Liss Rhod PE to facilitate fluorescence microscopy imaging. The resulting vesicles exhibited pronounced membrane fluctuations and non-spherical shapes, indicative of low membrane tension ($\sim \mathrm{nN/m}$) and excess membrane area \cite{van2024role}.

Adhesion between vesicles was induced by suspending them in a solution containing polyacrylamide (0.25 wt.-\%), a non-adsorbing polymer that gives rise to depletion-driven adhesion (see Materials and Methods). This effect occurs because polyacrylamide forms an exclusion layer near the membrane. Overlap of these layers between two GUVs reduces the excluded volume, resulting in an effective attractive interaction. This mechanism, previously applied to study particle engulfment by GUVs \cite{dinsmore1998hard,spanke2020wrapping,van2024role}, yields an adhesion energy $E_\mathrm{ad} = n \Delta V k_\mathrm{B} T$ \cite{asakura1954interaction}, where $n$ is the polymer number density and $\Delta V = A_\mathrm{c} d$ is the excluded volume reduction, with $A_\mathrm{c}$ as the contact area, and $d$ the interaction range. This expression simplifies to $E_\mathrm{ad} = w A_\mathrm{c}$, where $w = ndk_\mathrm{B}T$ is the effective adhesion strength per unit area.

\begin{figure}[t]
    \centering
    \includegraphics[width=1\linewidth]{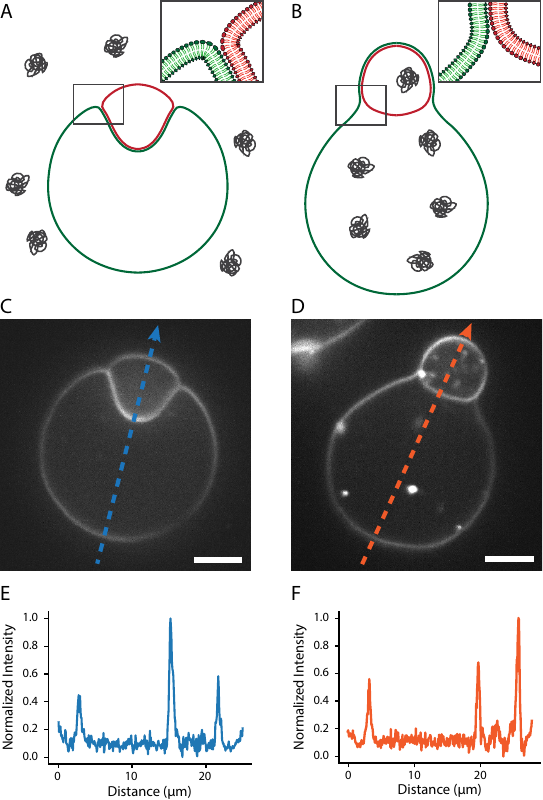}
    \caption{A,B) Schematic representation of endocytic (A) and exocytic (B) engulfment of a small vesicle (red membrane) by a large vesicle (green membrane) where the polymer is added to the outer and inner solution, respectively. C,D) Confocal fluorescence microscopy images of the mid-plane of partial endocytic (C) and exocytic (D) vesicle engulfment. E,F) Fluorescence intensity profile along the dashed lines from panel C and D. Scale bars are \qty{5}{\um}.}
    \label{fig:schematic}
\end{figure}

The direction of engulfment can be influenced by adding the polymer to either the outer or inner solution of the GUVs, as schematically depicted in Fig.~\ref{fig:schematic}A,B. When polymer is added to the outer solution, it creates depletion layers outside the GUVs, promoting adhesion to external vesicles and leading to endocytic engulfment (Fig.~\ref{fig:schematic}C). Conversely, adding polymer to the inner solution generates depletion layers inside the GUVs, which can drive adhesion to internal vesicles and lead to exocytic engulfment (Fig.~\ref{fig:schematic}D). 

The direction and degree of engulfment are reflected in the fluorescence signal of the GUV membranes (Fig.~\ref{fig:schematic}C,D). This is further confirmed by the intensity line profile through the centres of both vesicles, which shows three distinct peaks, corresponding to the free and adhered membranes of the large and small vesicle (Fig.~\ref{fig:schematic}E,F). In the adhered region, where the membranes overlap, the fluorescence intensity is elevated compared to that of a single membrane. For endocytic engulfment, the intensity profile shows an elevated central peak (Fig.~\ref{fig:schematic}E), whereas for exocytic engulfment, this shifts to one of the outer peaks (Fig.~\ref{fig:schematic}F). The degree of engulfment is quantified by the fraction of the small vesicle's membrane that exhibits higher fluorescence intensity.

\subsection*{Modelling the vesicle-vesicle engulfment}
To interpret the observed engulfment morphologies and their dependence on vesicle geometry, we modelled vesicle-vesicle interactions using energy minimization within a continuum framework (see Materials and Methods). To reflect experimental conditions, we assumed constant vesicle volume and membrane area. In experiments, osmotic conditions constrain GUV volume, while the high area compressibility modulus of lipid membranes (\qty{200}{\milli\newton\per\metre}) ensures negligible area changes under the weak adhesion interactions studied here ($O(10^2) \, k_\mathrm{B}T/\unit{\um^2}$). Under these constraints, the vesicle tensions and pressures become Lagrange multipliers, and the resulting interaction is governed by a balance between adhesion energy gain and bending energy cost. Increased contact area between the vesicles increases the total adhesion energy, while the associated membrane deformation incurs a bending energy penalty.

The model captures the experimental behaviour using four dimensionless parameters: the volume ratio of the small and large vesicle, their reduced volumes, and the ratio of the small vesicle size to the bendocapillary length. The volume ratio $\phi$ characterizes the relative size of the vesicles, and is defined as the volume of the small vesicle, $V_\mathrm{small}$, divided by that of the large vesicle, $V_\mathrm{large}$:
\begin{equation}
    \phi = \frac{V_\mathrm{small}}{V_\mathrm{large}}.
\end{equation}

\begin{figure}[t!]
    \centering
    \includegraphics[width=1.0\linewidth]{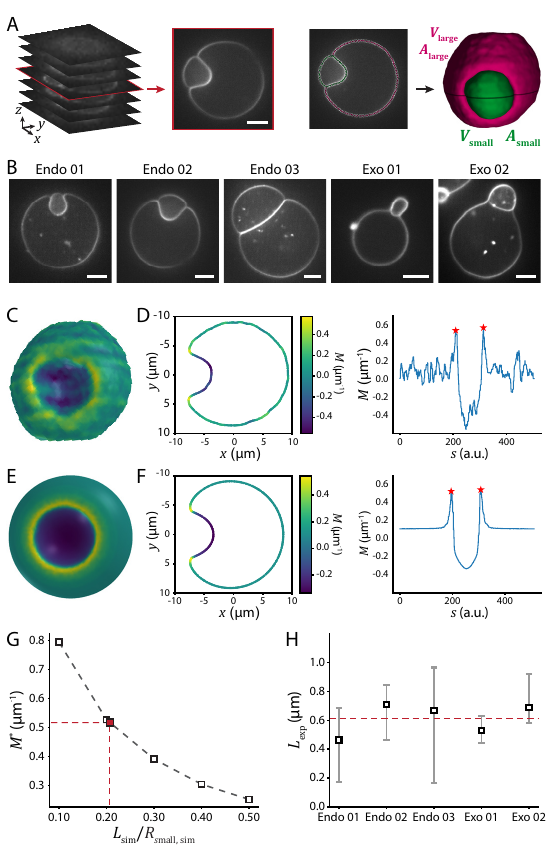}
    \caption{Calibration of $L_\mathrm{exp}$ through curvature matching. A) 3D reconstruction of vesicles from confocal fluorescence \textit{z}-stacks. Image stacks were segmented using the LimeSeg plugin in FIJI (ImageJ) \cite{machado2019limeseg,schindelin2012fiji}, enabling quantitative extraction of vesicle volume ($V$), surface area ($A$), mean curvature ($M$), and shape. B) Vesicle pairs that were used for the calibration of $L_\mathrm{exp}$. C-F) Peak mean curvature extraction for the experimentally measured  (\textit{Endo 02}) (C,D) and simulated (E,F) vesicle at $L_\mathrm{sim} = 0.2$. C) and E) show the 3D mesh with the colour coded mean curvature $M$. D) and F) show a single cross section as a contour plot (left panel) and line  plot (right panel). The peak values of $M$ are marked with a red star. G) Linear interpolation of $M^*_\mathrm{exp}$ (red square) to $M^*_\mathrm{sim}$ (open squares) as a function of $L_\mathrm{sim}/R_\mathrm{small,\,sim}$ for \textit{Endo 02}. H) $L_\mathrm{exp}$ for the 5 analysed vesicles (open squares) and their corresponding error bar, representing the standard deviation, which was determined from the standard deviation in $M^*_\mathrm{exp}$. The horizontal red dashed line represents the average $L_\mathrm{exp}$ value of the 5 measurements points. The scale bars are \qty{5}{\um}.}
    \label{fig:curv_main}
\end{figure}

The reduced volume $\nu_i$ ($i=$ small, large) quantifies the vesicle deformability, and is defined as the ratio between the vesicle's volume $V_i$ and the volume $V_{i,\mathrm{sph}}$ of a sphere having the same surface area $A_i$:
\begin{equation}
\nu_i = \frac{V_i}{V_{i, \mathrm{sph}}} = \frac{3\sqrt{4\pi} V_i}{A_i^{3/2}}.
\end{equation}
A vesicle with $\nu=1$ has no excess membrane area and adopts a spherical shape, behaving like a rigid particle. As $\nu$ decreases below 1, the amount of excess membrane area increases, allowing the vesicle to adopt a broader range of non-spherical shapes. The further $\nu$ deviates from 1, the more deformable the vesicle becomes, as more membrane area is available for shape transformations \cite{seifert1991shape}.

Finally, the ratio of the small vesicle size to the bendocapillary length, $R_\mathrm{small}/L$, characterizes the balance between adhesion and membrane bending forces, where $R_\mathrm{small}$ is given by 
\begin{equation}
    R_\mathrm{small} = \left(\frac{3}{4\pi}V_\mathrm{small}\right)^{1/3},
\end{equation}  
and $L$ is defined as
\begin{equation}
    L = \sqrt{\frac{\kappa}{w}},
    \label{eq:L}
\end{equation}  
with $\kappa$ the bending rigidity of the membrane. For $R_\mathrm{small}/L \gg 1$, adhesion and surface tension forces dominate over bending deformation forces, while for $R_\mathrm{small}/L \ll 1$, bending rigidity dominates.

The volume ratio $\phi$, and the reduced volumes of the large and small vesicle, $\nu_\mathrm{large}$ and $\nu_\mathrm{small}$, were determined experimentally by measuring the volume and surface area of each vesicle. These measurements were extracted from confocal microscopy \textit{z}-stacks using a contour segmentation algorithm, as illustrated in Fig.~\ref{fig:curv_main}A (see Materials and Methods). 

The experimental bendocapillary length, $L_\mathrm{exp}$, is constant for a given polymer concentration and can, in principle, be estimated from Eq.~\ref{eq:L}. Assuming a bending rigidity of $\kappa \approx 25\,k_\mathrm{B}T$, typical for DOPC membranes \cite{faizi2020fluctuation}, and estimating the adhesion strength as $w = 2 R_\mathrm{G} n k_\mathrm{B} T \approx 180 \, k_\mathrm{B}T/\qty{}{\um\squared}$, this yields a theoretical value of $L_\mathrm{exp} \approx \qty{0.37}{\um}$ (for 0.25 wt.\% polyacrylamide with a radius of gyration $R_\mathrm{G} \approx \qty{45}{\nm}$ \cite{francois1979polyacrylamide}). However, this estimate assumes contact adhesion between membranes and does not account for effects such as polymer flexibility \cite{tuinier2001excluded}, polydispersity, and membrane fluctuations \cite{helfrich1984undulations,lipowsky1995structure}, all of which can lower the effective adhesion strength.

Therefore, to obtain a more accurate estimate of $L_\mathrm{exp}$, we compared the curvature profiles of five experimentally observed vesicles (see Fig.~\ref{fig:curv_main}B) with those generated in simulations using varying values of the simulated bendocapillary length $L_\mathrm{sim}$. We expect the best agreement when the ratio $R_\mathrm{small}/L$ matches between experiment and simulation, which yields the condition:
\begin{equation}
    L_\mathrm{exp} = \frac{R_\mathrm{small,\, exp}}{R_\mathrm{small,\, sim}} \cdot L_\mathrm{sim}.
    \label{eq:lsim}
\end{equation}

For each experimental vesicle pair, we measured the peak mean curvature $M^*$ of the large vesicle, which typically occurs near the contact line and exhibits rotational symmetry around the \textit{z}-axis, defined by the line connecting the vesicles' centres of mass (see Supporting Fig.~S1). The procedure for extracting $M^*$ from both experimental and simulated vesicles is shown in Fig.~\ref{fig:curv_main}C-F and detailed in the Materials and Methods. We measured $M^*$ for five vesicle pairs spanning a range of morphologies and wrapping fractions (Fig.~\ref{fig:curv_main}B) and compared the experimental results to simulations (Fig.~\ref{fig:curv_main}G). Applying Eq.~\ref{eq:lsim} and averaging the results yielded $L_\mathrm{exp} = 0.61 \pm 0.10\, \unit{\micro\metre}$, as shown in Fig.~\ref{fig:curv_main}H.

Although $L_\mathrm{exp}$ should be approximately constant for a given polymer concentration, the vesicle size $R_\mathrm{small,\, exp}$ varies, causing changes in $R_\mathrm{small,\, exp}/L_\mathrm{exp}$. To account for this, we replicate the ratio $R_\mathrm{small}/L$ in simulations for each vesicle pair, enabling direct comparison between experiments and model predictions. This framework allows us to systematically investigate the role of the four dimensionless parameters in vesicle engulfment: the volume ratio ($\phi$), the reduced volumes of the small and large vesicle ($\nu_\mathrm{small}$ and $\nu_\mathrm{large}$), and the relative size with respect to the bendocapillary length ($R_\mathrm{small}/L$).

\begin{figure*}[!t]
    \centering
    \includegraphics[width=0.95\linewidth]{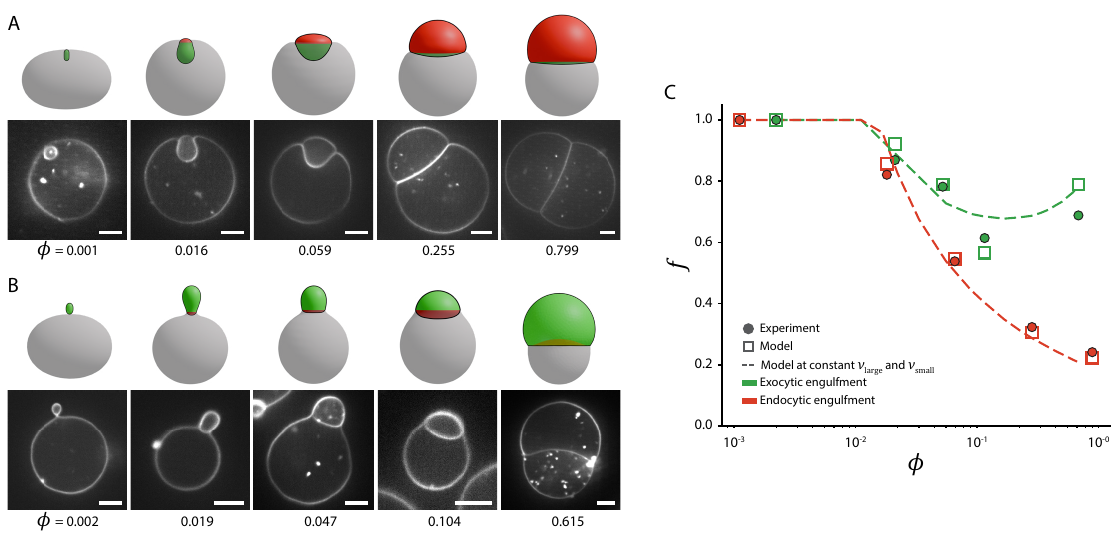}
    \caption{Endo- (A) and exocytic (B) vesicle engulfment as a function of volume ratio $\phi$. Confocal fluorescence microscopy images representing equilibrium conformations at various $\phi$ (indicated below the images). All vesicles had a reduced volume between 0.87 and 0.97 and were recorded at the same adhesion strength (0.25 wt.-\% polyacrylamide). The ratio $R_\mathrm{small}/L$ ranges from 1.6 to 19. For endocytic engulfment (A), the polymer was present in the outer solution, whereas for exocytic engulfment (B), the polymer was present in the inner solution. The corresponding morphologies, replicated by the model, are shown above the microscopy images. The large vesicle is represented in grey, the small vesicle in red, and the contact area is highlighted in green.
    C) Experimentally measured wrapping fractions (circles) and their simulated counterparts (squares) as a function of $\phi$ for the vesicles in panel A (red) and B (green).The dashed lines represent the simulated wrapping fraction for $\nu_\mathrm{large} = 0.9$ and $\nu_\mathrm{small} = 0.94$ for endocytic and $\nu_\mathrm{large} = 0.92$ and $\nu_\mathrm{small} = 0.92$ for exocytic engulfment. The ratio $R_\mathrm{small}/L = 5$ is held constant for the dashed lines.}
    \label{fig:SizeRatio}
\end{figure*}

\subsection*{Endo- and exocytic engulfment of vesicles with varying volume ratio}
Combining experiments and continuum simulations, we begin by studying how the direction of engulfment, whether endocytic or exocytic, influences vesicle morphology and the degree of engulfment (Fig.~\ref{fig:SizeRatio}). We explore this behaviour by studying vesicles pairs across a range of volume ratios ($\phi = 0.001 - 0.799$), while keeping the reduced volumes approximately constant. Due to experimental constraints, we restricted our analysis to vesicle pairs with reduced volumes between 0.87 and 0.97 for both large and small vesicles. We employ the same polymer concentration of 0.25 wt.-\% polyacrylamide, which should lead to a constant adhesion strength. However, due to vesicle size variation, the ratio $R_\mathrm{small}/L$ ranges from 1.6 to 19. 

The direction of engulfment is determined by where the polymer is added. If present in the outer solution, it promotes endocytic engulfment (Fig.~\ref{fig:SizeRatio}A); if added to the inner solution, it results in exocytic engulfment (Fig.~\ref{fig:SizeRatio}B). Since exocytic engulfment involves a GUV wrapping around an internal vesicle, it requires a configuration in which a large GUV encapsulates a smaller one (an example without adhesion is shown in Supporting Fig.~S2). Although such vesicle-in-vesicle configurations are less common with our droplet transfer method, they can be reliably generated using microfluidic approaches based on double emulsion techniques \cite{deng2017microfluidic}.

Due to the membranes’ flexibility, both vesicles undergo large deformations, mutually remodelling to minimize the system’s energy. The experimentally observed vesicle shapes are closely reproduced by our theoretical model for both endocytic and exocytic engulfment (Fig.~\ref{fig:SizeRatio}A,B), supporting its validity. In endocytic engulfment, the large vesicle forms an inward invagination at the contact site that deepens with decreasing $\phi$. Conversely, in exocytic engulfment, it forms an outward protrusion around the smaller vesicle. In both cases, the small vesicle undergoes pronounced shape changes as it becomes increasingly wrapped, transitioning from an oblate form in the shallow-wrapped state to a prolate, pear-like geometry in the deeply wrapped state. As the wrapping fraction approaches 1, a narrow catenoidal membrane neck forms, creating either an inward-facing bud (endocytic) or an outward-facing bud (exocytic) that fully encapsulates the smaller vesicle.
 
We note that some discrepancies between experiments and simulations can be seen for the smallest volume ratio (left-most panels of Fig.~\ref{fig:SizeRatio}A,B). In experiments, the large vesicle adopts a flattened, pancake-like shape with a circular cross-section, whereas simulations predict a more elongated, prolate form. This discrepancy can be attributed to gravitational effects that are present in experiments, caused by a density mismatch between the inner and outer sugar solutions (see Materials and Methods). To quantify these gravitational effects, we estimate the dimensionless gravity parameter $g = g_0 \Delta\rho R_\mathrm{large}^4 / \kappa$ \cite{kraus1995gravity}, where $g_0 = \qty{9.81}{\metre\per\second\squared}$ is the gravitational acceleration, $\Delta\rho \approx \qty{30}{\kilo\gram\per\metre\cubed}$ denotes the density difference between the inner (\qty{500}{\milli\Molar} sucrose) and outer solution (\qty{500}{\milli\Molar} glucose), and $\kappa \approx 25\,k_\mathrm{B}T$. The large vesicles of the fully wrapped endocytic and exocytic state have radii $R_\mathrm{large}$ of \qty{10.4}{\um} and \qty{8.6}{\um}, respectively, such that we find $g \geq 16$ for $R_\mathrm{large} \geq \qty{8.6}{\um}$. This value is in line with the predicted vesicle shapes in Ref.~\citealp{kraus1995gravity}, thus confirming that the shape discrepancy is a size-dependent gravitational effect.

\begin{figure*}[t]
    \centering
    \includegraphics[width=0.9\linewidth]{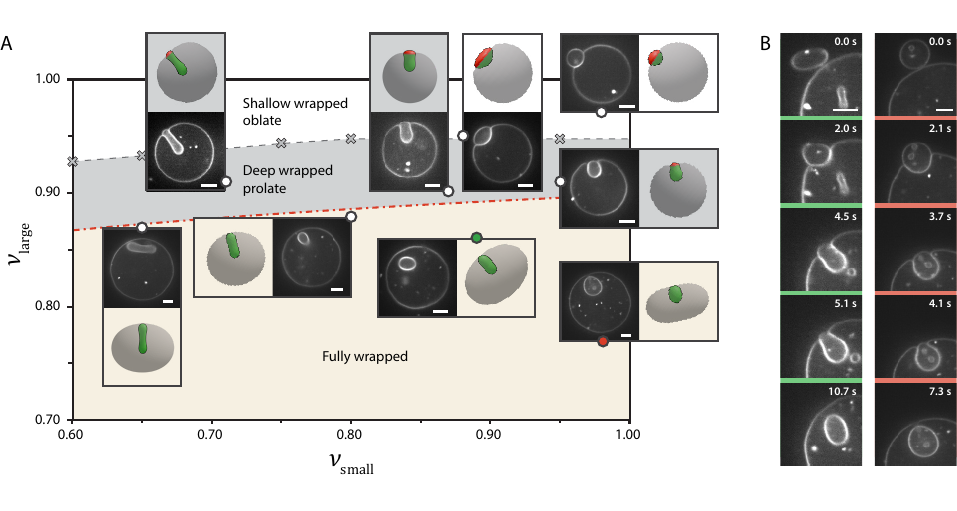}
    \caption{State diagram of endocytic vesicle engulfment as a function of $\nu_\mathrm{large}$ and $\nu_\mathrm{small}$. A) Vesicle-vesicle morphologies as a function of $\nu_\mathrm{large}$ (\textit{y}-axis) and $\nu_\mathrm{small}$ (\textit{x}-axis). Confocal fluorescence microscopy images represent typical equilibrium conformations of vesicle pairs for various reduced volumes. The corresponding morphologies that were replicated by the model are shown alongside the microscopy images. All vesicle pairs had a volume ratio between 0.013 and 0.037 and were recorded at the same adhesion strength (0.25 wt.-\% polyacrylamide). The ratio $R_\mathrm{small}/L$ ranges from 2.9 to 7.1. The shaded areas represent shallow wrapped oblate vesicles (white), deep wrapped prolate vesicles (grey), and fully wrapped vesicles (sand). The transition between the shallow and deep wrapped states (grey dashed line) was obtained from simulations (grey crosses). The transition to the fully wrapped state (red dash-dotted line) was calculated with Eq.~\ref{eq:nugamma} using $\phi = 0.016$.  
    B) Time series captured with confocal fluorescence microscopy, illustrating the morphology of two vesicles (green: $\nu_\mathrm{small}=0.89$ and red: $\nu_\mathrm{small}=0.98$ from panel A) as they transition from the free ($f=0$) to the fully engulfed state ($f=1$). Scale bars are \qty{5}{\um}.}
    \label{fig:RedVolume}
\end{figure*}

We quantified the degree of wrapping in both endo- and exocytic engulfment (Fig.~\ref{fig:SizeRatio}C) by measuring the wrapping fraction $f = A_\mathrm{c}/A_\mathrm{small}$ (see Supporting Text Section 1 and Supporting Fig.~S3 for details). For clarity, two sets of results are presented for the simulations. The open square symbols in Fig.~\ref{fig:SizeRatio}C provide the $f$ values when the four dimensionless parameters are matched with experiments for each vesicle pair. The dashed lines provide the $f$ values for varying $\phi$ when we use the representative values $\nu_\mathrm{large} = 0.9$, $\nu_\mathrm{small} = 0.94$, and $R_\mathrm{small}/L = 5$ for endocytic engulfment and $\nu_\mathrm{large} = 0.92$, $\nu_\mathrm{small} = 0.92$, and $R_\mathrm{small}/L = 5$ for exocytic engulfment. Overall, the experimental values of $f$ closely match simulation results across a broad range of volume ratios, supporting the accuracy of our model. For endocytic engulfment, $f$ increases monotonically as $\phi$ decreases, illustrating a continuous transition from shallow to full wrapping. In contrast, exocytic engulfment shows a non-monotonic trend as $f = 1$ for equal sized vesicles ($\phi = 1$), then decreases to a minimum near $\phi = 0.2$, before rising again toward full wrapping at lower $\phi$.

\subsection*{Endocytic vesicle engulfment and the role of vesicle reduced volume}
While our system enables the study of both endo- and exocytic engulfment as shown in Fig.~\ref{fig:SizeRatio}, we now focus on endocytic engulfment to further explore how vesicle deformability influences wrapping morphology, building on our findings regarding engulfment direction and volume ratio. To this end, we map vesicle-vesicle morphologies as a function $\nu_\mathrm{small}$ and $\nu_\mathrm{large}$, while maintaining an approximately constant volume ratio $\phi$ (Fig.~\ref{fig:RedVolume}A). The reduced volume of the small vesicle determines its deformability: even slight deviations from a fully inflated state ($\nu_\mathrm{small} = 1$) allow the vesicle to deform, enabling it to adapt to the curvature imposed by the large vesicle. In the absence of interactions, vesicles with reduced volumes between 0.7 and 1.0 typically adopt either prolate or spherical equilibrium shapes \cite{seifert1991shape}. However, when a vesicle becomes (partially) engulfed by another, their interaction significantly alters the minimum energy shape of the system. Conversely, the reduced volume of the large vesicle reflects the amount of excess membrane area available for wrapping. At fixed volume ratio, this parameter governs the extent of engulfment: as $\nu_\mathrm{large}$ decreases from 1, the wrapping state transitions smoothly from shallow, to deep, and finally to full engulfment. To illustrate this, Fig.~\ref{fig:RedVolume}A shows experimental and simulated vesicle morphologies, delineating the shallow, deep, and fully wrapped regimes.

In the partially wrapped state, the large vesicle remains mostly spherical with an invagination at the contact point, while the shape of the small vesicle is influenced by both $\nu_\mathrm{small}$ and $\nu_\mathrm{large}$. In the shallow wrapped state, the small vesicle adopts an oblate form, which varies from pancake-like to nearly spherical as $\nu_\mathrm{small}$ increases. In the deep wrapped state, the small vesicle becomes prolate, elongating as $\nu_\mathrm{small}$ decreases (Fig.~\ref{fig:RedVolume}A).  This dynamic between wrapping and reduced volume reveals how $\nu_\mathrm{large}$ governs the small vesicle's morphology during the interaction.

In the fully wrapped state, the small vesicle attains the morphology that it would have in the absence of interactions, with the key difference being that it is now enclosed by the membrane of the large vesicle. The equilibrium shape of the large vesicle differs slightly between experiments and simulations. As discussed previously, this can be attributed to a size-dependent gravitational effect \cite{kraus1995gravity}. Indeed, all vesicles that exhibit a deviation have a radius $R_\mathrm{large}$ that is larger than \qty{10}{\um}.

\begin{figure*}[t]
    \centering
    \includegraphics[width=0.9\linewidth]{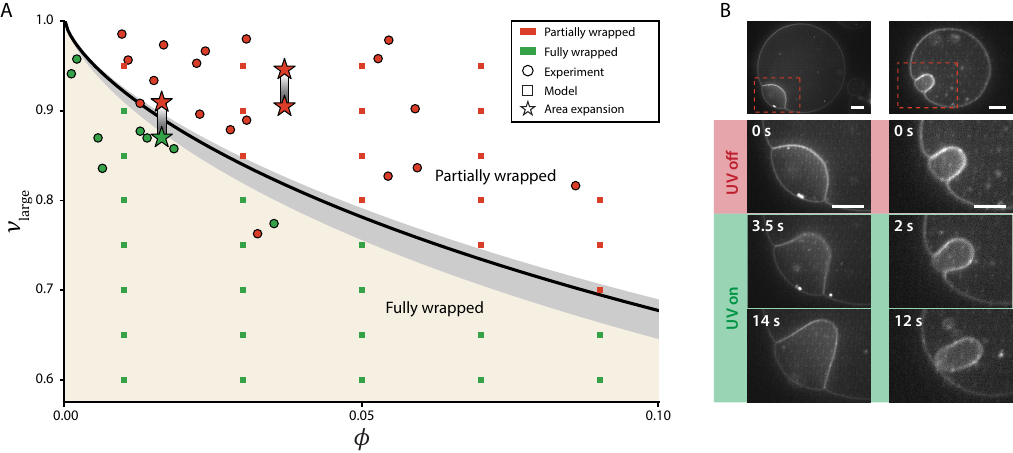}
    \caption{A) State diagram of the partially and fully wrapped state as a function of $\nu_\mathrm{large}$ and $\phi$. The black line gives the theoretical transition between the partially (red) and fully wrapped (green) state based on Eq.~\ref{eq:nugamma}. The grey shaded area indicates the influence of $\nu_\mathrm{small}$ on this transition. Based on the available experimental data, we show a lower and an upper bound of $\nu_\mathrm{small} = 0.7$ and $1.0$, respectively, and the black line representing $\nu_\mathrm{small} = 0.9$. Experimental and simulated data points are plotted as circles and squares, respectively. For the simulated data, $R_\mathrm{small}/L = 10$ and $\nu_\mathrm{small} = 0.9$, while for the experimental data $R_\mathrm{small}/L$ ranges from 1.7 to 19 and $\nu_\mathrm{small}$ from 0.65 to 0.99. Area expansion measurements corresponding to panel B are given by pairs of stars, where the top star corresponds to the vesicle pair before UV illumination, and the bottom star to the vesicle pair after UV illumination. 
    B) Confocal fluorescence microscopy images of two vesicles containing azo-PC before (red) and during (green) UV exposure. The increase in membrane area during UV exposure drives a transition from shallow to deep wrapped (left column) and from deep to fully wrapped state (right column). Scale bars are \qty{10}{\um} and \qty{5}{\um} for the left and right column, respectively.}
    \label{fig:StateDiagram}
\end{figure*}

The dynamic transition of a small vesicle from the free to fully wrapped state is shown in Fig.~\ref{fig:RedVolume}B and Supporting Movies S2-S3 for two different values of $\nu_\mathrm{small}$, corresponding to the vesicle pairs in Fig.~\ref{fig:RedVolume}A marked with green ($\nu_\mathrm{small}=0.89$) and red ($\nu_\mathrm{small}=0.98$). In both cases, the small vesicle progressively adapts its shape as the degree of wrapping increases. Upon initial contact, it begins to bulge toward the large vesicle, inducing an invagination in the latter. As wrapping deepens, a neck forms at the contact line and gradually narrows in diameter. At this stage, the small vesicle assumes a pear-like shape, with its narrow end positioned at the neck. The neck continues to constrict until the small vesicle is fully wrapped. In the final state, the neck becomes too small to resolve with a microscope, yet it still connects the inward bud with the membrane of the large vesicle. The small vesicle regains its pre-wrapping shape after full engulfment. Although the specific shapes differ based on $\nu_\mathrm{small}$, the general steps of this engulfment process remain consistent.

\subsection*{Geometrically constrained engulfment}
To further characterize the onset of full engulfment, we examine how this transition depends on vesicle geometry. In the $R_\mathrm{small}/L>>1$ regime, the full engulfment transition is determined only by the system's geometry. Whether a vesicle will become completely wrapped can thus be predicted based on $\phi$, $\nu_\mathrm{large}$, and $\nu_\mathrm{small}$. When wrapping the small vesicle, the large vesicle requires part of its excess membrane area to encapsulate it, becoming more spherical. Additionally, the remaining membrane area has to envelop both the original volume of the large vesicle, as well as the volume of the small vesicle. The effective reduced volume of the large vesicle after wrapping a small vesicle is thus \cite{satarifard2023mutual,bahrami2014wrapping}:

\begin{align}
    \nu_\gamma = 3 \sqrt{4\pi} \frac{V_\mathrm{large} + V_\mathrm{small}}{(A_\mathrm{large} - A_\mathrm{small})^{3/2}},
\end{align}
which can be rewritten in terms of the dimensionless parameters as (see Supporting Text Section 2):
\begin{align}
    \nu_\gamma = \frac{1 + \phi}{\left( \nu_{\mathrm{large}}^{-2/3} - \phi^{2/3}\nu_{\mathrm{small}}^{-2/3}\right)^{3/2}}.
    \label{eq:nugamma}
\end{align}

The value of $\nu_\gamma$ quantifies whether or not full wrapping is possible given the available excess membrane area. When $\nu_\gamma \leq 1$, the small vesicle can be fully engulfed. In contrast, when $\nu_\gamma > 1$, the large vesicle lacks sufficient excess membrane area to fully engulf the small vesicle, resulting in only partial wrapping. Interestingly, Eq.~\ref{eq:nugamma} shows that the engulfment transition depends only weakly on $\nu_\mathrm{small}$, since it is scaled by $\phi^{2/3}$, which  is small in our study ($\phi < 0.1$). This weak dependence highlights that, within the $R_\mathrm{small}/L \gg 1$ regime, the small vesicle deformability plays a limited role in the wrapping outcome.

We constructed a state diagram mapping the wrapping outcome as a function of $\phi$ and $\nu_\mathrm{large}$ (Fig.~\ref{fig:StateDiagram}A), combining experimental and simulation data with the theoretical transition from Eq.~\ref{eq:nugamma}. The weak dependence on $\nu_\mathrm{small}$ is illustrated by the narrow shaded gray region in Fig.~\ref{fig:StateDiagram}A, which represents the predicted transition boundary for $\nu_\mathrm{small}$ ranging from 0.7 to 1.0. Both the experimental data (with the exception of one data point close to the transition boundary) and the model accurately reflect this transition. Notably, the experimental data spans a wide range of $R_\mathrm{small}/L$ values, from $R_\mathrm{small}/L = 1.9$ to 19, yet still fall along the predicted transition boundary. This suggests that even modest increases beyond $R_\mathrm{small}/L = 1$ are sufficient to reach the geometry-dominated regime. For instance, simulations at $R_\mathrm{small}/L = 10$ already conform well to the analytical prediction. Therefore, we conclude that the onset of geometry-governed wrapping behaviour occurs at relatively small values of $R_\mathrm{small}/L$.

In this regime, Eq.~\ref{eq:nugamma} and Fig.~\ref{fig:StateDiagram}A suggest that decreasing $\nu_\mathrm{large}$ can drive the transition of a small vesicle from partial to full wrapping. To experimentally test this, we incorporated photo-responsive azo-PC lipids into the GUV membranes \cite{pernpeintner2017light} (Materials and Methods). Upon UV illumination (365 nm), azo-PC undergoes trans-to-cis photoisomerization, which increases membrane area and thereby reduces $\nu_\mathrm{large}$ \cite{aleksanyan2023photomanipulation}. 
Using this light-triggered area modulation, we actively controlled the degree of vesicle wrapping, inducing transitions from shallow to deep, and from deep to fully wrapped states (Fig.~\ref{fig:StateDiagram} and Supporting Movies S4–S5). These results demonstrate that modulating membrane area, and consequently $\nu_\mathrm{large}$, offers a powerful mechanism to control engulfment, consistent with recent studies on photoswitchable endocytosis of biomolecular condensates \cite{mangiarotti2024photoswitchable}.

While the majority of the experimental data conform to the predicted transition, one data point near $\phi \approx 0.035$ remains only partially wrapped despite satisfying $\nu_\gamma < 1$. Comparison with a neighbouring point that does reach full engulfment, despite having nearly identical parameters, reveals a key difference in their $R_\mathrm{small}/L$ value: $R_\mathrm{small}/L = 7.1$ for the fully wrapped vesicle, versus 2.9 for the partially wrapped one. This comparison suggests that the outlier lies too close to the boundary of the geometry-dominated regime, and that additional factors such as bending energy may inhibit full engulfment when $R_\mathrm{small}/L$ is closer to unity.

\subsection*{Adhesion constrained engulfment}
\begin{figure}[t]
    \centering
    \includegraphics[width=1\linewidth]{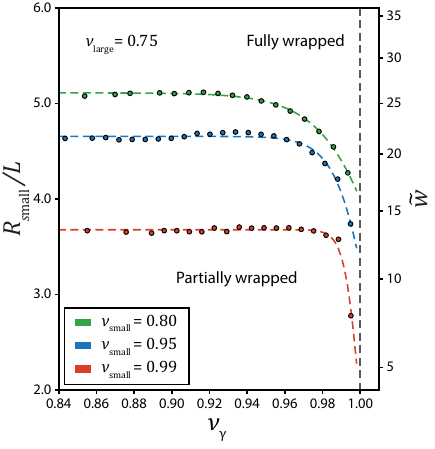}
    \caption{The partial to full wrapping transition as a function of $R_\mathrm{small}/L$ (left axis), or equivalently $\tilde{w} = w R_\mathrm{small}^2/\kappa$ (right axis), and $\nu_\gamma$. The transition is shown for constant $\nu_\mathrm{large} = 0.75$, while $\nu_\mathrm{small}$ is varied: 0.8 (green), 0.95 (blue) and 0.99 (red). The vertical grey dashed line indicates the geometric cut-off at $\nu_{\gamma}=1$, beyond which full engulfment cannot occur.}
    \label{fig:RoverL}
\end{figure}

Having established that vesicle geometry primarily dictates the engulfment transition in the $R_\mathrm{small}/L \gg 1$ regime, we now turn to the regime where $R_\mathrm{small}/L$ approaches unity. In this limit, geometric effects alone are no longer sufficient to predict the wrapping outcome. Although full engulfment can still occur when $\nu_{\gamma} < 1$, the increased interplay between bending rigidity and adhesion introduces additional constraints. As a result, full wrapping may be suppressed despite sufficient excess membrane area, and the deformability of the small vesicle, quantified by its reduced volume, becomes a critical factor in determining the engulfment transition.

In Fig.~\ref{fig:RoverL}, we show the full engulfment transition as a function of $R_\mathrm{small}/L$ (or equivalently $\tilde{w} = w R_\mathrm{small}^2/\kappa$) and the effective reduced volume of the large vesicle, $\nu_\mathrm{\gamma}$  (see Eq.~\ref{eq:nugamma}). We fix $\nu_\mathrm{large} = 0.75$, and examine three values of $\nu_\mathrm{small}$: 0.8, 0.95, and 0.99. The variation in $\nu_\mathrm{\gamma}$, shown on the \textit{x}-axis of the phase diagram of Fig.~\ref{fig:RoverL}, is obtained by changing the volume ratio $\phi$. The partial to full wrapping transition is calculated from curves that describe how the total energy varies with the wrapping fraction, as detailed in Supporting Fig.~S4. The resulting transitions show that full engulfment requires increasing adhesion strength (or increasing $R_\mathrm{small}/L$) as $\nu_\mathrm{small}$ decreases, illustrating that increased deformability of the small vesicle inhibits complete wrapping. Notably, the two experimental points discussed previously, both at $\nu_\gamma = 0.93$ but with $R_\mathrm{small}/L$ values of 7.1 and 2.9, fall into the fully and partially wrapped regimes, respectively, consistent with their observed wrapping states.

For all three values of $\nu_\mathrm{small}$, the transition curves follow a similar trend: a plateau at low $\nu_\gamma$, followed by a steep decline as $\nu_\gamma$ approaches unity. Since we assume there is no spontaneous curvature in our system, the bending energy is scale-independent. Therefore, it depends only on the overall shape of the vesicles, not on their size. Consequently, the total bending energy pre-engulfment is the same irrespective of the volume ratio $\phi$ and hence $\nu_\gamma$. During engulfment, as $\nu_\mathrm{\gamma}$ increases toward unity, the large vesicle transitions from a prolate to a near-spherical shape. This shape change reduces the bending energy cost associated with encapsulating the small vesicle, facilitating full engulfment. This can be observed in the results shown in Supporting Fig.~S4 and explains the decreasing adhesion strength required for full wrapping. The effect is strongest near $\nu_\gamma = 1$, where the bending energy gain is largest. A similar mechanism has been described in theoretical studies of nanoparticle engulfment by vesicles~\cite{bahrami2014wrapping}. We can also compare this effect for different values of $\nu_\mathrm{small}$. As shown in Supporting Fig.~S4, the wrapping energy curves for $\nu_\mathrm{small} = 0.8$ remain relatively similar across varying $\nu_\gamma$, resulting in a gradual decrease in the required adhesion as $\nu_{\gamma}$ approaches unity. In contrast, for $\nu_\mathrm{small} = 0.99$, the energy curves vary significantly with $\nu_\gamma$, leading to a sharper drop in the adhesion threshold for complete wrapping.

\section*{Discussion}
In this study, we developed a biomimetic model system, consisting of large GUVs engulfing smaller ones, to investigate how vesicle deformability influences vesicle–vesicle engulfment. This interaction was driven by membrane adhesion through depletion interactions, providing a simplified but biologically relevant analogue of vesicle internalization by the cell membrane. By selectively placing the depletant in either the inner or outer solution of the GUVs, we were able to influence the directionality of engulfment. Remarkably, this allowed us to observe both endocytic and exocytic engulfment pathways within the same system.

We simulated the experimental system using a discretized continuum model where the balance between bending and adhesion energy contributions is characterised by the bendocapillary length, $L$. Benchmarking experiments against simulations we found $L_\mathrm{exp} = 0.61 \pm 0.10\, \unit{\micro\metre}$, for which the model accurately reproduced vesicle morphologies and wrapping fractions across the full range of experimental conditions. The strong agreement between simulations and experiments indicates that the vesicle engulfment is governed by the interplay between membrane bending and adhesion energies, the only contributions included in the model. This energy balance gives rise to two distinct engulfment regimes, which are determined by the dimensionless ratio $R_\mathrm{small}/L$.

In the regime where $R_\mathrm{small}/L>1$, the wrapping fraction is governed by geometric constraints. In this limit, whether the large vesicle partially or fully engulfs the smaller one depends on the amount of excess membrane available for wrapping. The transition to full engulfment can therefore be described by the system’s geometry: the volume ratio $\phi$, the reduced volume of the large vesicle $\nu_\mathrm{large}$, and, to a lesser extent, the reduced volume of the small vesicle $\nu_\mathrm{small}$. Similar geometry-driven wrapping transitions have been predicted theoretically for both spherical particles and biomolecular condensates \cite{bahrami2014wrapping,satarifard2023mutual}. We experimentally demonstrated that geometry governs the wrapping process by controlling the degree of wrapping through area modulation of azo-PC doped vesicles.

Interestingly, while $\nu_\mathrm{small}$ has little effect on the full engulfment transition, it plays a significant role in shaping the minimum-energy configurations of partially wrapped states. While shallow wrapped, the small vesicle adopts an oblate shape, lying flat on the large vesicle's surface with minimal indentation. In contrast, in deep wrapping, the small vesicle transitions to a prolate shape, oriented perpendicular to the large vesicle's membrane, penetrating into it. These behaviours effectively combine the stability of shallow-wrapped oblates with that of deep-wrapped prolates, as predicted for rigid ellipsoids \cite{bahrami2013orientational,agudo2020engulfment}. However, unlike rigid ellipsoids or rods, which reorient during the wrapping transition \cite{agudo2020engulfment,bahrami2013orientational,van2024role,dasgupta2014shape}, the deformable vesicles in our system reshape themselves, transitioning between oblate and prolate regimes as the wrapping fraction increases \cite{midya2023membrane,yi2016incorporation}.

In the regime where $R_\mathrm{small}/L \approx 1$, adhesion is no longer the only dominant factor, and the full engulfment transition becomes sensitive to the deformability of the small vesicle, characterized by its reduced volume $\nu_\mathrm{small}$. Lower values of $\nu_\mathrm{small}$, corresponding to higher deformability, inhibit full engulfment, even when it is geometrically feasible for the large vesicle to wrap the smaller one. Consequently, the adhesion strength required for full engulfment increases significantly as $\nu_\mathrm{small}$ decreases. This behaviour is consistent with prior findings for planar membranes and volume-unconstrained vesicles \cite{midya2023membrane,yi2016incorporation}. For $\nu_{\gamma}$ values slightly below unity, where the large vesicle has just enough excess membrane to potentially achieve full engulfment, the required adhesion energy for full engulfment decreases. This counter-intuitive effect was previously predicted for rigid spherical particles \cite{bahrami2014wrapping}, and arises from an energetic relaxation of the large vesicle.

Our system serves as a biomimetic model for the engulfment of deformable particles, such as vesicles, viral capsids, and drug delivery carriers, by cellular membranes. By highlighting how vesicle deformability and the bendocapillary length influence morphology and engulfment transitions, our results advance the understanding of membrane-mediated uptake mechanisms. These insights offer design principles for engineering more effective drug delivery systems based on soft, deformable carriers. In particular, by tuning the mechanical properties of the transport vesicle, the efficiency and selectivity of cargo delivery could be actively controlled through membrane-mediated engulfment.


\newpage
\section*{Materials and methods}

\section*{Experimental}

\subsection*{Materials}
Unless otherwise specified, all chemicals were used as received. The lipids, 1,2-dioleoyl-sn-glycero-3-phosphocholine (DOPC) and fluorescent 1,2-dioleoyl-sn-glycero-3-phosphoethanolamine-N-(lissamine rhodamine B sulfonyl) (ammonium salt) (Liss Rhod PE) in chloroform, and 1-stearoyl-2-[(E)-4-(4-((4-butylphenyl)diazenyl)phenyl)butanoyl]-sn-glycero-3-phosphocholine (azo-PC) in powder form, were obtained from Avanti Polar Lipids (Alabaster, AL). Chloroform ($\geq$99.5\%), heavy mineral oil, glucose, and sucrose were purchased from Sigma Aldrich. A solution of 10 wt.-\% polyacrylamide (molecular weight: \qtyrange{700000}{1000000}{\g/\mol}) in water was obtained from Polysciences Inc. All aqueous solutions were prepared using ultrapure Milli-Q\textsuperscript{\textregistered} water. Sugar solutions were filtered through a \qty{0.2}{\um} cellulose filter (VWR).

Lipid stock solutions were prepared by dissolving or diluting the lipids in chloroform to the following final concentrations: \qty{12}{\mg/\ml} and \qty{3}{\mg/\ml} for DOPC, \qty{0.2}{\mg/\ml} for Liss Rhod PE, and \qty{3}{\mg/\ml} for azo-PC. All stock solutions were stored at \qty{-20}{\degreeCelsius} until use.

Microscopy measurements were conducted using custom observation chambers fabricated from \qtyproduct{24 x 50}{\mm} \#1.5 cover glasses, purchased from VWR. The chambers were assembled by cutting off \qty{0.5}{\cm} of the base of a \qty{1}{\ml} pipette tip and gluing it onto a cover glass using UV-curable glue (Norland Optical Adhesive 68). The glue was cured by placing the chamber under a \qty{365}{\nm} UV LED (Nichia NVSU233B SMD LED UV, \qty{365}{\nm}, \qty{1450}{\mW}) for several minutes. During measurements, the chamber was sealed with Parafilm\textsuperscript{\textregistered}.

\subsection*{Preparation of GUVs}
Giant unilamellar vesicles (GUVs) were prepared using a modified droplet transfer method based on Refs.~\citealp{vutukuri2020active,van2024role}. A lipid-in-oil solution (LOS) was prepared by mixing \qty{100}{\ul} of DOPC (\qty{12}{\mg/\ml}) and \qty{10}{\ul} of Liss Rhod PE (\qty{0.2}{\mg/\ml}) stock in a \qty{20}{\ml} glass vial. 
The chloroform was evaporated under a gentle stream of N\textsubscript{2} while rotating the vial to form a thin film of lipids at the bottom. The film was kept under vacuum in a desiccator for 1-2 \unit{\hour} to remove residual chloroform. Next, \qty{3}{\g} of heavy mineral oil was added to the vial, resulting in a lipid concentration of \qty{425}{\micro\Molar} (\qty{0.1}{\mol\percent} Liss Rhod PE). The mixture was sonicated for \qty{1}{\hour} at \qty{40}{\degreeCelsius}. Finally, to ensure proper dissolution, the LOS was kept overnight in the dark at room temperature. 

For the incorporation of azo-PC into the GUVs, the same protocol was followed with two modifications. The lipid composition was adjusted to \qty{80}{\ul} of DOPC (\qty{3}{\mg/\ml}), \qty{3}{\ul} of Liss Rhod PE (\qty{0.2}{\mg/\ml}), and \qty{27}{\ul} of azo-PC (\qty{3}{\mg/\ml}). Additionally, the quantity of mineral oil was reduced to \qty{1.5}{\g}, resulting in a lipid concentration of \qty{225}{\micro\Molar} (\qty{0.1}{\mol\percent} Liss Rhod PE and \qty{24}{\mol\percent} azo-PC).

To prepare the GUVs, \qty{200}{\ul} of LOS was gently layered on top of \qty{500}{\ul} of the outer solution (\qty{500}{\milli\Molar} glucose) in a \qty{2}{\ml} Eppendorf\textsuperscript{\textregistered} tube. In a second \qty{2}{\ml} Eppendorf\textsuperscript{\textregistered} tube, \qty{200}{\ul} of LOS and \qty{15}{\ul} of the inner solution were combined. The inner solution was either \qty{500}{\milli\Molar} sucrose or \qty{500}{\milli\Molar} sucrose with 0.25 wt.-\% polyacrylamide, depending on the desired adhesion properties. The solution without polyacrylamide was used when studying endocytic engulfment, while the solution with polyacrylamide was used when studying exocytic engulfment, as it induced adhesion on the inside of the GUVs. The mixture was mechanically agitated on an Eppendorf\textsuperscript{\textregistered} tube rack for \qty{10}{\s} to form an emulsion of water-in-oil droplets \cite{moga2019optimization}. Next, \qty{120}{\ul} of the emulsion was gently layered on top of the water-oil column in the first Eppendorf\textsuperscript{\textregistered} tube and immediately centrifuged at \qty{200}{\g} for \qty{2}{\min}. After centrifugation, the top oil layer was carefully removed with a pipette, and the remaining GUV solution was left to sediment for \qtyrange{30}{60}{\min} before being used for measurements.

\subsection*{GUV-polymer suspensions and sample preparation}
To study endocytic engulfment, \qty{20}{\ul} of the GUV solution, taken from the bottom of the tube, was gently mixed with \qty{20}{\ul} of 0.5 wt.-\% polyacrylamide in \qty{500}{\milli\Molar} glucose by pipetting the solution up and down several times, resulting in a final concentration of 0.25 wt.-\% polyacrylamide in the outer solution. To investigate exocytic engulfment, where the inner solution of the GUVs contained 0.25 wt.-\% polyacrylamide, the GUV solution was instead mixed with \qty{20}{\ul} of \qty{500}{\milli\Molar} glucose without polyacrylamide. Subsequently, \qty{10}{\ul} of either mixture was added to a custom imaging chamber and sealed with Parafilm\textsuperscript{\textregistered} to prevent evaporation. A new chamber was used for each experiment.

\subsection*{Confocal microscopy imaging}
Microscopy measurements were conducted using a confocal laser scanning microscope (Nikon Eclipse Ti-U inverted microscope with a VTinfinity3 CLSM module, Visitech), equipped with a Hamamatsu ORCA-Flash4.0 CMOS camera and a 100x, 1.49 NA oil objective lens. Liss Rhod PE labelled GUVs were excited with a \qty{561}{\nm} laser. Time-lapse imaging was performed at a frame rate of 10 fps. \textit{Z}-scans were conducted with a step size of \qty{0.25}{\um} and an exposure time of \qty{50}{\ms}, resulting in a scan rate of approximately 8 slices per second.

\subsection*{Area expansion of azo-PC lipids}
Membrane area expansion in GUVs containing \qty{25}{\mol\percent} azo-PC was achieved by illuminating the sample with an external UV LED (Nichia NVSU233B SMD LED UV, \qty{365}{\nm}, \qty{1450}{\mW}) \cite{aleksanyan2023photomanipulation,frank2016photoswitchable,pernpeintner2017light,mangiarotti2024photoswitchable}. The UV LED was positioned above the condenser lens of the confocal microscope and focused using an additional lens mounted on a custom optical rail. A low pass filter was added to the UV light path to filter out higher wavelengths. Illumination with UV light resulted in increased GUV membrane area, evidenced by significant membrane fluctuations and shape deformations, indicating a trans-to-cis transition of the azo-PC lipids \cite{aleksanyan2023photomanipulation}. However, simultaneous imaging with the \qty{561}{\nm} laser suppressed this expansion. Turning off the UV LED while continuing illumination with the \qty{561}{\nm} laser induced the opposite transition (cis-to-trans), as shown by rapid GUV membrane shrinkage. To minimize the influence of the \qty{561}{\nm} laser and maximize area expansion, time-lapse imaging was performed at low laser power with a frame rate of \qty{500}{\ms} per frame.

We estimate the membrane area increase to be approximately 3\% based on comparison with other vesicle-vesicle morphologies. This is lower than expected given the molar ratio of azo-PC and area expansions reported in literature \cite{aleksanyan2023photomanipulation}. We suspect that the droplet transfer method causes the fraction of azo-PC incorporated in the GUVs to be lower than intended \cite{weakly2024several}. Nonetheless, the light modulation effectively demonstrates the significance of excess membrane area, and allows for control over the wrapping configuration.

\subsection*{Analysis} 
The volume and surface area of the GUVs were determined from confocal \textit{z}-scans using the LimeSeg plugin in Fiji (ImageJ) \cite{schindelin2012fiji,machado2019limeseg}. LimeSeg, a particle-based active contour method, segments 3D objects by detecting their outlines, such as those of fluorescently labelled GUVs.

The \textit{z}-spacing of the \textit{z}-stacks was corrected by a factor of 0.83 to account for spherical aberration caused by imaging in an aqueous medium with an oil objective lens. This correction factor, obtained from Ref.~\citealp{diel2020tutorial}, was calculated using the corresponding ImageJ plugin with the following parameters: numerical aperture (NA) = 1.49, imaging medium refractive index = 1.33, and immersion oil refractive index = 1.52. The accuracy of the correction factor was validated using spherical GUVs (see Supporting Fig.~S5).

\subsection*{Calibration of $L_\mathrm{exp}$ using curvature matching}
To determine $L_\mathrm{exp}$, we compared the membrane curvature of five experimental vesicle pairs with simulated vesicles generated using varying values of $L_\mathrm{sim}$. Confocal fluorescence \textit{z}-stacks of each vesicle pair were imported into FIJI (ImageJ) and segmented into 3D meshes using the LimeSeg plugin \cite{machado2019limeseg,schindelin2012fiji} (see Fig.~\ref{fig:curv_main}A). The resulting meshes were exported from FIJI as point clouds and processed with a custom Python script, which reconstructed the 3D surface mesh using the Open3D library.

Each vesicle pair was then replicated in simulations using the experimentally measured dimensionless parameters: $\phi$, $\nu_\mathrm{large}$, and $\nu_\mathrm{small}$ as input, while $L_\mathrm{sim}$ was treated as a tuneable parameter. The simulated vesicle shapes were converted into point clouds and processed using the same in-house Python script as for the experimental data. To ensure consistency in mesh resolution and eliminate potential biases, the simulated meshes were downsampled using a voxel-based approach to match the resolution of the experimental meshes.

The peak mean curvature $M^*$ was extracted from both experimental and simulated meshes using an identical analysis. First, the local mean curvature $M = \frac{1}{2} \left(c_1 + c_2 \right)$, where $c_1 = r_1^{-1}$ and $c_2 = r_2^{-1}$ are the principal curvatures, was computed at each vertex of the large vesicle mesh via quadric surface fitting over the local vertex neighbourhood using the libigl library \cite{libigl}. Representative visualizations of the 3D mesh and the corresponding colour-coded curvature map for vesicle \textit{Endo 02} are shown in Fig.~\ref{fig:curv_main}C,E.
Then, 9 cross-sections were extracted by intersecting the mesh with planes rotated about the \textit{z}-axis (see Supporting Fig.~S1) in \ang{20} increments. From each cross-section, the curvature profile along the intersection curve was computed. The two local peaks in each profile were extracted (marked with red stars in Fig.~\ref{fig:curv_main}D,F), yielding 18 curvature values per vesicle. These values were averaged to obtain a single representative curvature measurement for the experimental vesicle, $M^*_\mathrm{exp}$. The same procedure was applied to the simulated meshes to compute a series of $M^*_\mathrm{sim}$ values corresponding to different values of $L_\mathrm{sim}$.

For each vesicle pair, the value of $M^*_\mathrm{exp}$ was compared to the simulated values $M^*_\mathrm{sim}(L_\mathrm{sim})$. The value of $L_\mathrm{sim}$ that best matched the experimental data was identified by linear interpolation (Fig.~\ref{fig:curv_main}G). Finally, to obtain $L_\mathrm{exp}$, $L_\mathrm{sim}$ was multiplied by $R_\mathrm{small,\,exp}$ (Fig.~\ref{fig:curv_main}H). In Supporting Fig.~S1, slice comparisons between the experimentally measured vesicles and those simulated using Surface Evolver are shown.

\section*{Numerical}
\subsection*{Free energy}

For the simulations, we use a variation of the Helfrich free energy model for both the large and small vesicles. The total energy of the system is given by,

\begin{equation}
\begin{aligned}
E=&  \sum_{\substack{i=\mathrm{small},\\ \mathrm{large}}} \left( 
\int_{A_{i}} \mathrm{d} S_i \left[2 \kappa M^2 \right] 
+ \sigma_{i} A_{i}
+ P_{i} V_{i} \right)
\\&-  w \int_{A_{\mathrm{c}}} \mathrm{d} S,
\end{aligned}
\label{eq:energyfunctional1}
\end{equation}
where $A_{i}$ and $V_{i}$ ($i =$ small, large) represent the membrane area and the volume of the small and large vesicles. The first term in the summation accounts for the curvature energy, where $\kappa$ is the bending rigidity and $M$ the mean curvature of the membrane. We assume zero spontaneous curvature. The other two terms ensure a constant membrane area and volume, where $\sigma_{i}$ and $P_{i}$ are the Lagrange multipliers for the membrane area and the volume constraints, respectively. The final term represents the adhesion energy, with $w$ denoting the adhesion strength due to depletion interaction and $A_\mathrm{c}$ the contact area between the vesicles.

In the case of full wrapping, the membrane of the bud that encapsulates the small vesicle remains connected to the large vesicle through a narrow catenoidal membrane neck. Since resolving the neck region is expensive computationally and the catenoidal neck has zero bending energy, we have used an approximation where we model the encapsulated small vesicle and the remainder of the large vesicle as two separate bodies, without explicitly modelling the catenoid-shape neck. In the Supporting Information, we justify this approximation quantitatively (see Supporting Fig.~S6).

\subsection*{Energy minimisation}

In order to find the minimum free energy configurations, we employ the software package Surface Evolver (SE) package version 2.72k. The membrane surfaces of both the large and small vesicles are discretised using a triangulated mesh.

We calculate the mean curvature using the Surface Evolver Method `star\_perp\_sq\_mean\_curvature'. We specify which facets belong to which vesicle and calculate the mean curvature at each vertex as

\begin{equation}
\begin{aligned}
M_v = \frac{3}{2}\frac{\nabla A_v \cdot \nabla V_v}{\nabla A_v \cdot \nabla A_v},
\end{aligned}
\label{eq:energyfunctional2}
\end{equation}
where $\nabla A_v$ is the area gradient at a vertex $v$ and $\nabla V_v$ is the volume gradient at a vertex $v$. Then, the total bending energy of the system is given by

\begin{equation}
\begin{aligned}
E_b = 2 \sum_{\substack{i=\mathrm{small},\\ \mathrm{large}}} \sum_{v} \frac{A_{iv} \kappa_i}{3} M^2_{iv},
\end{aligned}
\label{eq:energyfunctional3}
\end{equation}
where we have summed over each vertex on each vesicle. We assume both vesicles have equal membrane bending constants, $\kappa_\mathrm{small} = \kappa_\mathrm{large} = \kappa$. 
The areas of the vesicle are held constant by using the facet area method in Surface Evolver, which employs the tension as a Lagrange multiplier. We also use a level constraint of $z=0$ on the three-phase contact line to restrict the rotational and translational degrees of freedom.

\section*{Acknowledgements}
S.v.d.H and H.R.V. thank Frieder Mugele and Mireille Claessens for kindly providing access to confocal and fluorescence microscopes. H.R.V. acknowledges funding from The Netherlands Organization for Scientific Research (NWO, Dutch Science Foundation). H.K. acknowledges funding from EPSRC EP/V034154/2. A.W.B. acknowledges funding from EPSRC CDT on Molecular Sciences for Medicine EP/S022791/1.

\section*{Author contributions}
H.R.V. and S.v.d.H. initiated the project; S.v.d.H. performed experiments; A.B. performed simulations; S.v.d.H. and A.B. analysed data; all authors contributed to the research design and in writing the manuscript.

\section*{Competing interests}
The authors declare that they have no competing interests.

\section*{Data and materials availability}
All data needed to evaluate the conclusions in the paper are present in the paper and/or the Supporting Information.

\newpage
\bibliography{Bibliography}  

\begin{thebibliography}{51}%
\makeatletter
\providecommand \@ifxundefined [1]{%
 \@ifx{#1\undefined}
}%
\providecommand \@ifnum [1]{%
 \ifnum #1\expandafter \@firstoftwo
 \else \expandafter \@secondoftwo
 \fi
}%
\providecommand \@ifx [1]{%
 \ifx #1\expandafter \@firstoftwo
 \else \expandafter \@secondoftwo
 \fi
}%
\providecommand \natexlab [1]{#1}%
\providecommand \enquote  [1]{``#1''}%
\providecommand \bibnamefont  [1]{#1}%
\providecommand \bibfnamefont [1]{#1}%
\providecommand \citenamefont [1]{#1}%
\providecommand \href@noop [0]{\@secondoftwo}%
\providecommand \href [0]{\begingroup \@sanitize@url \@href}%
\providecommand \@href[1]{\@@startlink{#1}\@@href}%
\providecommand \@@href[1]{\endgroup#1\@@endlink}%
\providecommand \@sanitize@url [0]{\catcode `\\12\catcode `\$12\catcode `\&12\catcode `\#12\catcode `\^12\catcode `\_12\catcode `\%12\relax}%
\providecommand \@@startlink[1]{}%
\providecommand \@@endlink[0]{}%
\providecommand \url  [0]{\begingroup\@sanitize@url \@url }%
\providecommand \@url [1]{\endgroup\@href {#1}{\urlprefix }}%
\providecommand \urlprefix  [0]{URL }%
\providecommand \Eprint [0]{\href }%
\providecommand \doibase [0]{https://doi.org/}%
\providecommand \selectlanguage [0]{\@gobble}%
\providecommand \bibinfo  [0]{\@secondoftwo}%
\providecommand \bibfield  [0]{\@secondoftwo}%
\providecommand \translation [1]{[#1]}%
\providecommand \BibitemOpen [0]{}%
\providecommand \bibitemStop [0]{}%
\providecommand \bibitemNoStop [0]{.\EOS\space}%
\providecommand \EOS [0]{\spacefactor3000\relax}%
\providecommand \BibitemShut  [1]{\csname bibitem#1\endcsname}%
\let\auto@bib@innerbib\@empty
\bibitem [{\citenamefont {Gy{\"o}rgy}\ \emph {et~al.}(2011)\citenamefont {Gy{\"o}rgy}, \citenamefont {Szab{\'o}}, \citenamefont {P{\'a}szt{\'o}i}, \citenamefont {P{\'a}l}, \citenamefont {Misj{\'a}k}, \citenamefont {Aradi}, \citenamefont {L{\'a}szl{\'o}}, \citenamefont {P{\'a}llinger}, \citenamefont {Pap}, \citenamefont {Kittel} \emph {et~al.}}]{gyorgy2011membrane}%
  \BibitemOpen
  \bibfield  {author} {\bibinfo {author} {\bibfnamefont {B.}~\bibnamefont {Gy{\"o}rgy}}, \bibinfo {author} {\bibfnamefont {T.~G.}\ \bibnamefont {Szab{\'o}}}, \bibinfo {author} {\bibfnamefont {M.}~\bibnamefont {P{\'a}szt{\'o}i}}, \bibinfo {author} {\bibfnamefont {Z.}~\bibnamefont {P{\'a}l}}, \bibinfo {author} {\bibfnamefont {P.}~\bibnamefont {Misj{\'a}k}}, \bibinfo {author} {\bibfnamefont {B.}~\bibnamefont {Aradi}}, \bibinfo {author} {\bibfnamefont {V.}~\bibnamefont {L{\'a}szl{\'o}}}, \bibinfo {author} {\bibfnamefont {E.}~\bibnamefont {P{\'a}llinger}}, \bibinfo {author} {\bibfnamefont {E.}~\bibnamefont {Pap}}, \bibinfo {author} {\bibfnamefont {A.}~\bibnamefont {Kittel}}, \emph {et~al.},\ }\bibfield  {title} {\bibinfo {title} {Membrane vesicles, current state-of-the-art: emerging role of extracellular vesicles},\ }\href@noop {} {\bibfield  {journal} {\bibinfo  {journal} {Cellular and Molecular Life Sciences}\ }\textbf {\bibinfo {volume} {68}},\ \bibinfo {pages} {2667} (\bibinfo {year} {2011})}\BibitemShut
  {NoStop}%
\bibitem [{\citenamefont {Arandjelovic}\ and\ \citenamefont {Ravichandran}(2015)}]{arandjelovic2015phagocytosis}%
  \BibitemOpen
  \bibfield  {author} {\bibinfo {author} {\bibfnamefont {S.}~\bibnamefont {Arandjelovic}}\ and\ \bibinfo {author} {\bibfnamefont {K.~S.}\ \bibnamefont {Ravichandran}},\ }\bibfield  {title} {\bibinfo {title} {Phagocytosis of apoptotic cells in homeostasis},\ }\href@noop {} {\bibfield  {journal} {\bibinfo  {journal} {Nature Immunology}\ }\textbf {\bibinfo {volume} {16}},\ \bibinfo {pages} {907} (\bibinfo {year} {2015})}\BibitemShut {NoStop}%
\bibitem [{\citenamefont {Cohen}(2016)}]{cohen2016viruses}%
  \BibitemOpen
  \bibfield  {author} {\bibinfo {author} {\bibfnamefont {F.~S.}\ \bibnamefont {Cohen}},\ }\bibfield  {title} {\bibinfo {title} {How viruses invade cells},\ }\href@noop {} {\bibfield  {journal} {\bibinfo  {journal} {Biophysical Journal}\ }\textbf {\bibinfo {volume} {110}},\ \bibinfo {pages} {1028} (\bibinfo {year} {2016})}\BibitemShut {NoStop}%
\bibitem [{\citenamefont {Marsh}\ and\ \citenamefont {Helenius}(2006)}]{marsh2006virus}%
  \BibitemOpen
  \bibfield  {author} {\bibinfo {author} {\bibfnamefont {M.}~\bibnamefont {Marsh}}\ and\ \bibinfo {author} {\bibfnamefont {A.}~\bibnamefont {Helenius}},\ }\bibfield  {title} {\bibinfo {title} {Virus entry: open sesame},\ }\href@noop {} {\bibfield  {journal} {\bibinfo  {journal} {Cell}\ }\textbf {\bibinfo {volume} {124}},\ \bibinfo {pages} {729} (\bibinfo {year} {2006})}\BibitemShut {NoStop}%
\bibitem [{\citenamefont {Klein}\ \emph {et~al.}(2020)\citenamefont {Klein}, \citenamefont {Cortese}, \citenamefont {Winter}, \citenamefont {Wachsmuth-Melm}, \citenamefont {Neufeldt}, \citenamefont {Cerikan}, \citenamefont {Stanifer}, \citenamefont {Boulant}, \citenamefont {Bartenschlager},\ and\ \citenamefont {Chlanda}}]{klein2020sars}%
  \BibitemOpen
  \bibfield  {author} {\bibinfo {author} {\bibfnamefont {S.}~\bibnamefont {Klein}}, \bibinfo {author} {\bibfnamefont {M.}~\bibnamefont {Cortese}}, \bibinfo {author} {\bibfnamefont {S.~L.}\ \bibnamefont {Winter}}, \bibinfo {author} {\bibfnamefont {M.}~\bibnamefont {Wachsmuth-Melm}}, \bibinfo {author} {\bibfnamefont {C.~J.}\ \bibnamefont {Neufeldt}}, \bibinfo {author} {\bibfnamefont {B.}~\bibnamefont {Cerikan}}, \bibinfo {author} {\bibfnamefont {M.~L.}\ \bibnamefont {Stanifer}}, \bibinfo {author} {\bibfnamefont {S.}~\bibnamefont {Boulant}}, \bibinfo {author} {\bibfnamefont {R.}~\bibnamefont {Bartenschlager}},\ and\ \bibinfo {author} {\bibfnamefont {P.}~\bibnamefont {Chlanda}},\ }\bibfield  {title} {\bibinfo {title} {Sars-cov-2 structure and replication characterized by in situ cryo-electron tomography},\ }\href@noop {} {\bibfield  {journal} {\bibinfo  {journal} {Nature Communications}\ }\textbf {\bibinfo {volume} {11}},\ \bibinfo {pages} {5885} (\bibinfo {year} {2020})}\BibitemShut {NoStop}%
\bibitem [{\citenamefont {Bayati}\ \emph {et~al.}(2021)\citenamefont {Bayati}, \citenamefont {Kumar}, \citenamefont {Francis},\ and\ \citenamefont {McPherson}}]{bayati2021sars}%
  \BibitemOpen
  \bibfield  {author} {\bibinfo {author} {\bibfnamefont {A.}~\bibnamefont {Bayati}}, \bibinfo {author} {\bibfnamefont {R.}~\bibnamefont {Kumar}}, \bibinfo {author} {\bibfnamefont {V.}~\bibnamefont {Francis}},\ and\ \bibinfo {author} {\bibfnamefont {P.~S.}\ \bibnamefont {McPherson}},\ }\bibfield  {title} {\bibinfo {title} {Sars-cov-2 infects cells after viral entry via clathrin-mediated endocytosis},\ }\href@noop {} {\bibfield  {journal} {\bibinfo  {journal} {Journal of Biological Chemistry}\ }\textbf {\bibinfo {volume} {296}} (\bibinfo {year} {2021})}\BibitemShut {NoStop}%
\bibitem [{\citenamefont {Allen}\ and\ \citenamefont {Cullis}(2013)}]{allen2013liposomal}%
  \BibitemOpen
  \bibfield  {author} {\bibinfo {author} {\bibfnamefont {T.~M.}\ \bibnamefont {Allen}}\ and\ \bibinfo {author} {\bibfnamefont {P.~R.}\ \bibnamefont {Cullis}},\ }\bibfield  {title} {\bibinfo {title} {Liposomal drug delivery systems: from concept to clinical applications},\ }\href@noop {} {\bibfield  {journal} {\bibinfo  {journal} {Advanced Drug Delivery Reviews}\ }\textbf {\bibinfo {volume} {65}},\ \bibinfo {pages} {36} (\bibinfo {year} {2013})}\BibitemShut {NoStop}%
\bibitem [{\citenamefont {Guimar{\~a}es}\ \emph {et~al.}(2021)\citenamefont {Guimar{\~a}es}, \citenamefont {Cavaco-Paulo},\ and\ \citenamefont {Nogueira}}]{guimaraes2021design}%
  \BibitemOpen
  \bibfield  {author} {\bibinfo {author} {\bibfnamefont {D.}~\bibnamefont {Guimar{\~a}es}}, \bibinfo {author} {\bibfnamefont {A.}~\bibnamefont {Cavaco-Paulo}},\ and\ \bibinfo {author} {\bibfnamefont {E.}~\bibnamefont {Nogueira}},\ }\bibfield  {title} {\bibinfo {title} {Design of liposomes as drug delivery system for therapeutic applications},\ }\href@noop {} {\bibfield  {journal} {\bibinfo  {journal} {International Journal of Pharmaceutics}\ }\textbf {\bibinfo {volume} {601}},\ \bibinfo {pages} {120571} (\bibinfo {year} {2021})}\BibitemShut {NoStop}%
\bibitem [{\citenamefont {Bareford}\ and\ \citenamefont {Swaan}(2007)}]{bareford2007endocytic}%
  \BibitemOpen
  \bibfield  {author} {\bibinfo {author} {\bibfnamefont {L.~M.}\ \bibnamefont {Bareford}}\ and\ \bibinfo {author} {\bibfnamefont {P.~W.}\ \bibnamefont {Swaan}},\ }\bibfield  {title} {\bibinfo {title} {Endocytic mechanisms for targeted drug delivery},\ }\href@noop {} {\bibfield  {journal} {\bibinfo  {journal} {Advanced Drug Delivery Reviews}\ }\textbf {\bibinfo {volume} {59}},\ \bibinfo {pages} {748} (\bibinfo {year} {2007})}\BibitemShut {NoStop}%
\bibitem [{\citenamefont {Joshi}\ \emph {et~al.}(2020)\citenamefont {Joshi}, \citenamefont {de~Beer}, \citenamefont {Giepmans},\ and\ \citenamefont {Zuhorn}}]{joshi2020endocytosis}%
  \BibitemOpen
  \bibfield  {author} {\bibinfo {author} {\bibfnamefont {B.~S.}\ \bibnamefont {Joshi}}, \bibinfo {author} {\bibfnamefont {M.~A.}\ \bibnamefont {de~Beer}}, \bibinfo {author} {\bibfnamefont {B.~N.}\ \bibnamefont {Giepmans}},\ and\ \bibinfo {author} {\bibfnamefont {I.~S.}\ \bibnamefont {Zuhorn}},\ }\bibfield  {title} {\bibinfo {title} {Endocytosis of extracellular vesicles and release of their cargo from endosomes},\ }\href@noop {} {\bibfield  {journal} {\bibinfo  {journal} {ACS Nano}\ }\textbf {\bibinfo {volume} {14}},\ \bibinfo {pages} {4444} (\bibinfo {year} {2020})}\BibitemShut {NoStop}%
\bibitem [{\citenamefont {Conner}\ and\ \citenamefont {Schmid}(2003)}]{conner2003regulated}%
  \BibitemOpen
  \bibfield  {author} {\bibinfo {author} {\bibfnamefont {S.~D.}\ \bibnamefont {Conner}}\ and\ \bibinfo {author} {\bibfnamefont {S.~L.}\ \bibnamefont {Schmid}},\ }\bibfield  {title} {\bibinfo {title} {Regulated portals of entry into the cell},\ }\href@noop {} {\bibfield  {journal} {\bibinfo  {journal} {Nature}\ }\textbf {\bibinfo {volume} {422}},\ \bibinfo {pages} {37} (\bibinfo {year} {2003})}\BibitemShut {NoStop}%
\bibitem [{\citenamefont {Lipowsky}\ and\ \citenamefont {D{\"o}bereiner}(1998)}]{lipowsky1998vesicles}%
  \BibitemOpen
  \bibfield  {author} {\bibinfo {author} {\bibfnamefont {R.}~\bibnamefont {Lipowsky}}\ and\ \bibinfo {author} {\bibfnamefont {H.-G.}\ \bibnamefont {D{\"o}bereiner}},\ }\bibfield  {title} {\bibinfo {title} {Vesicles in contact with nanoparticles and colloids},\ }\href@noop {} {\bibfield  {journal} {\bibinfo  {journal} {Europhysics Letters}\ }\textbf {\bibinfo {volume} {43}},\ \bibinfo {pages} {219} (\bibinfo {year} {1998})}\BibitemShut {NoStop}%
\bibitem [{\citenamefont {Dasgupta}\ \emph {et~al.}(2014)\citenamefont {Dasgupta}, \citenamefont {Auth},\ and\ \citenamefont {Gompper}}]{dasgupta2014shape}%
  \BibitemOpen
  \bibfield  {author} {\bibinfo {author} {\bibfnamefont {S.}~\bibnamefont {Dasgupta}}, \bibinfo {author} {\bibfnamefont {T.}~\bibnamefont {Auth}},\ and\ \bibinfo {author} {\bibfnamefont {G.}~\bibnamefont {Gompper}},\ }\bibfield  {title} {\bibinfo {title} {Shape and orientation matter for the cellular uptake of nonspherical particles},\ }\href@noop {} {\bibfield  {journal} {\bibinfo  {journal} {Nano Letters}\ }\textbf {\bibinfo {volume} {14}},\ \bibinfo {pages} {687} (\bibinfo {year} {2014})}\BibitemShut {NoStop}%
\bibitem [{\citenamefont {van~der Ham}\ \emph {et~al.}(2024)\citenamefont {van~der Ham}, \citenamefont {Agudo-Canalejo},\ and\ \citenamefont {Vutukuri}}]{van2024role}%
  \BibitemOpen
  \bibfield  {author} {\bibinfo {author} {\bibfnamefont {S.}~\bibnamefont {van~der Ham}}, \bibinfo {author} {\bibfnamefont {J.}~\bibnamefont {Agudo-Canalejo}},\ and\ \bibinfo {author} {\bibfnamefont {H.~R.}\ \bibnamefont {Vutukuri}},\ }\bibfield  {title} {\bibinfo {title} {Role of shape in particle-lipid membrane interactions: from surfing to full engulfment},\ }\href@noop {} {\bibfield  {journal} {\bibinfo  {journal} {ACS Nano}\ }\textbf {\bibinfo {volume} {18}},\ \bibinfo {pages} {10407} (\bibinfo {year} {2024})}\BibitemShut {NoStop}%
\bibitem [{\citenamefont {Bahrami}(2013)}]{bahrami2013orientational}%
  \BibitemOpen
  \bibfield  {author} {\bibinfo {author} {\bibfnamefont {A.~H.}\ \bibnamefont {Bahrami}},\ }\bibfield  {title} {\bibinfo {title} {Orientational changes and impaired internalization of ellipsoidal nanoparticles by vesicle membranes},\ }\href@noop {} {\bibfield  {journal} {\bibinfo  {journal} {Soft Matter}\ }\textbf {\bibinfo {volume} {9}},\ \bibinfo {pages} {8642} (\bibinfo {year} {2013})}\BibitemShut {NoStop}%
\bibitem [{\citenamefont {Agudo-Canalejo}\ and\ \citenamefont {Lipowsky}(2015)}]{agudo2015critical}%
  \BibitemOpen
  \bibfield  {author} {\bibinfo {author} {\bibfnamefont {J.}~\bibnamefont {Agudo-Canalejo}}\ and\ \bibinfo {author} {\bibfnamefont {R.}~\bibnamefont {Lipowsky}},\ }\bibfield  {title} {\bibinfo {title} {Critical particle sizes for the engulfment of nanoparticles by membranes and vesicles with bilayer asymmetry},\ }\href@noop {} {\bibfield  {journal} {\bibinfo  {journal} {ACS Nano}\ }\textbf {\bibinfo {volume} {9}},\ \bibinfo {pages} {3704} (\bibinfo {year} {2015})}\BibitemShut {NoStop}%
\bibitem [{\citenamefont {Agudo-Canalejo}(2021)}]{agudo2021particle}%
  \BibitemOpen
  \bibfield  {author} {\bibinfo {author} {\bibfnamefont {J.}~\bibnamefont {Agudo-Canalejo}},\ }\bibfield  {title} {\bibinfo {title} {Particle engulfment by strongly asymmetric membranes with area reservoirs},\ }\href@noop {} {\bibfield  {journal} {\bibinfo  {journal} {Soft Matter}\ }\textbf {\bibinfo {volume} {17}},\ \bibinfo {pages} {298} (\bibinfo {year} {2021})}\BibitemShut {NoStop}%
\bibitem [{\citenamefont {Spanke}\ \emph {et~al.}(2020)\citenamefont {Spanke}, \citenamefont {Style}, \citenamefont {Fran{\c{c}}ois-Martin}, \citenamefont {Feofilova}, \citenamefont {Eisentraut}, \citenamefont {Kress}, \citenamefont {Agudo-Canalejo},\ and\ \citenamefont {Dufresne}}]{spanke2020wrapping}%
  \BibitemOpen
  \bibfield  {author} {\bibinfo {author} {\bibfnamefont {H.~T.}\ \bibnamefont {Spanke}}, \bibinfo {author} {\bibfnamefont {R.~W.}\ \bibnamefont {Style}}, \bibinfo {author} {\bibfnamefont {C.}~\bibnamefont {Fran{\c{c}}ois-Martin}}, \bibinfo {author} {\bibfnamefont {M.}~\bibnamefont {Feofilova}}, \bibinfo {author} {\bibfnamefont {M.}~\bibnamefont {Eisentraut}}, \bibinfo {author} {\bibfnamefont {H.}~\bibnamefont {Kress}}, \bibinfo {author} {\bibfnamefont {J.}~\bibnamefont {Agudo-Canalejo}},\ and\ \bibinfo {author} {\bibfnamefont {E.~R.}\ \bibnamefont {Dufresne}},\ }\bibfield  {title} {\bibinfo {title} {Wrapping of microparticles by floppy lipid vesicles},\ }\href@noop {} {\bibfield  {journal} {\bibinfo  {journal} {Physical Review Letters}\ }\textbf {\bibinfo {volume} {125}},\ \bibinfo {pages} {198102} (\bibinfo {year} {2020})}\BibitemShut {NoStop}%
\bibitem [{\citenamefont {Bahrami}\ \emph {et~al.}(2014)\citenamefont {Bahrami}, \citenamefont {Raatz}, \citenamefont {Agudo-Canalejo}, \citenamefont {Michel}, \citenamefont {Curtis}, \citenamefont {Hall}, \citenamefont {Gradzielski}, \citenamefont {Lipowsky},\ and\ \citenamefont {Weikl}}]{bahrami2014wrapping}%
  \BibitemOpen
  \bibfield  {author} {\bibinfo {author} {\bibfnamefont {A.~H.}\ \bibnamefont {Bahrami}}, \bibinfo {author} {\bibfnamefont {M.}~\bibnamefont {Raatz}}, \bibinfo {author} {\bibfnamefont {J.}~\bibnamefont {Agudo-Canalejo}}, \bibinfo {author} {\bibfnamefont {R.}~\bibnamefont {Michel}}, \bibinfo {author} {\bibfnamefont {E.~M.}\ \bibnamefont {Curtis}}, \bibinfo {author} {\bibfnamefont {C.~K.}\ \bibnamefont {Hall}}, \bibinfo {author} {\bibfnamefont {M.}~\bibnamefont {Gradzielski}}, \bibinfo {author} {\bibfnamefont {R.}~\bibnamefont {Lipowsky}},\ and\ \bibinfo {author} {\bibfnamefont {T.~R.}\ \bibnamefont {Weikl}},\ }\bibfield  {title} {\bibinfo {title} {Wrapping of nanoparticles by membranes},\ }\href@noop {} {\bibfield  {journal} {\bibinfo  {journal} {Advances in Colloid and Interface Science}\ }\textbf {\bibinfo {volume} {208}},\ \bibinfo {pages} {214} (\bibinfo {year} {2014})}\BibitemShut {NoStop}%
\bibitem [{\citenamefont {Yi}\ \emph {et~al.}(2011)\citenamefont {Yi}, \citenamefont {Shi},\ and\ \citenamefont {Gao}}]{yi2011cellular}%
  \BibitemOpen
  \bibfield  {author} {\bibinfo {author} {\bibfnamefont {X.}~\bibnamefont {Yi}}, \bibinfo {author} {\bibfnamefont {X.}~\bibnamefont {Shi}},\ and\ \bibinfo {author} {\bibfnamefont {H.}~\bibnamefont {Gao}},\ }\bibfield  {title} {\bibinfo {title} {Cellular uptake of elastic nanoparticles},\ }\href@noop {} {\bibfield  {journal} {\bibinfo  {journal} {Physical Review Letters}\ }\textbf {\bibinfo {volume} {107}},\ \bibinfo {pages} {098101} (\bibinfo {year} {2011})}\BibitemShut {NoStop}%
\bibitem [{\citenamefont {Yi}\ and\ \citenamefont {Gao}(2016)}]{yi2016incorporation}%
  \BibitemOpen
  \bibfield  {author} {\bibinfo {author} {\bibfnamefont {X.}~\bibnamefont {Yi}}\ and\ \bibinfo {author} {\bibfnamefont {H.}~\bibnamefont {Gao}},\ }\bibfield  {title} {\bibinfo {title} {Incorporation of soft particles into lipid vesicles: Effects of particle size and elasticity},\ }\href@noop {} {\bibfield  {journal} {\bibinfo  {journal} {Langmuir}\ }\textbf {\bibinfo {volume} {32}},\ \bibinfo {pages} {13252} (\bibinfo {year} {2016})}\BibitemShut {NoStop}%
\bibitem [{\citenamefont {Imoto}\ \emph {et~al.}(2022)\citenamefont {Imoto}, \citenamefont {Raychaudhuri}, \citenamefont {Ma}, \citenamefont {Fenske}, \citenamefont {Sandoval}, \citenamefont {Itoh}, \citenamefont {Blumrich}, \citenamefont {Matsubayashi}, \citenamefont {Mamer}, \citenamefont {Zarebidaki} \emph {et~al.}}]{imoto2022dynamin}%
  \BibitemOpen
  \bibfield  {author} {\bibinfo {author} {\bibfnamefont {Y.}~\bibnamefont {Imoto}}, \bibinfo {author} {\bibfnamefont {S.}~\bibnamefont {Raychaudhuri}}, \bibinfo {author} {\bibfnamefont {Y.}~\bibnamefont {Ma}}, \bibinfo {author} {\bibfnamefont {P.}~\bibnamefont {Fenske}}, \bibinfo {author} {\bibfnamefont {E.}~\bibnamefont {Sandoval}}, \bibinfo {author} {\bibfnamefont {K.}~\bibnamefont {Itoh}}, \bibinfo {author} {\bibfnamefont {E.-M.}\ \bibnamefont {Blumrich}}, \bibinfo {author} {\bibfnamefont {H.~T.}\ \bibnamefont {Matsubayashi}}, \bibinfo {author} {\bibfnamefont {L.}~\bibnamefont {Mamer}}, \bibinfo {author} {\bibfnamefont {F.}~\bibnamefont {Zarebidaki}}, \emph {et~al.},\ }\bibfield  {title} {\bibinfo {title} {Dynamin is primed at endocytic sites for ultrafast endocytosis},\ }\href@noop {} {\bibfield  {journal} {\bibinfo  {journal} {Neuron}\ }\textbf {\bibinfo {volume} {110}},\ \bibinfo {pages} {2815} (\bibinfo {year} {2022})}\BibitemShut {NoStop}%
\bibitem [{\citenamefont {Tang}\ \emph {et~al.}(2016)\citenamefont {Tang}, \citenamefont {Zhang}, \citenamefont {Ye},\ and\ \citenamefont {Zheng}}]{tang2016wrapping}%
  \BibitemOpen
  \bibfield  {author} {\bibinfo {author} {\bibfnamefont {H.}~\bibnamefont {Tang}}, \bibinfo {author} {\bibfnamefont {H.}~\bibnamefont {Zhang}}, \bibinfo {author} {\bibfnamefont {H.}~\bibnamefont {Ye}},\ and\ \bibinfo {author} {\bibfnamefont {Y.}~\bibnamefont {Zheng}},\ }\bibfield  {title} {\bibinfo {title} {Wrapping of a deformable nanoparticle by the cell membrane: insights into the flexibility-regulated nanoparticle-membrane interaction},\ }\href@noop {} {\bibfield  {journal} {\bibinfo  {journal} {Journal of Applied Physics}\ }\textbf {\bibinfo {volume} {120}} (\bibinfo {year} {2016})}\BibitemShut {NoStop}%
\bibitem [{\citenamefont {Midya}\ \emph {et~al.}(2023)\citenamefont {Midya}, \citenamefont {Auth},\ and\ \citenamefont {Gompper}}]{midya2023membrane}%
  \BibitemOpen
  \bibfield  {author} {\bibinfo {author} {\bibfnamefont {J.}~\bibnamefont {Midya}}, \bibinfo {author} {\bibfnamefont {T.}~\bibnamefont {Auth}},\ and\ \bibinfo {author} {\bibfnamefont {G.}~\bibnamefont {Gompper}},\ }\bibfield  {title} {\bibinfo {title} {Membrane-mediated interactions between nonspherical elastic particles},\ }\href@noop {} {\bibfield  {journal} {\bibinfo  {journal} {ACS Nano}\ }\textbf {\bibinfo {volume} {17}},\ \bibinfo {pages} {1935} (\bibinfo {year} {2023})}\BibitemShut {NoStop}%
\bibitem [{\citenamefont {Satarifard}\ and\ \citenamefont {Lipowsky}(2023)}]{satarifard2023mutual}%
  \BibitemOpen
  \bibfield  {author} {\bibinfo {author} {\bibfnamefont {V.}~\bibnamefont {Satarifard}}\ and\ \bibinfo {author} {\bibfnamefont {R.}~\bibnamefont {Lipowsky}},\ }\bibfield  {title} {\bibinfo {title} {Mutual remodeling of interacting nanodroplets and vesicles},\ }\href@noop {} {\bibfield  {journal} {\bibinfo  {journal} {Communications Physics}\ }\textbf {\bibinfo {volume} {6}},\ \bibinfo {pages} {6} (\bibinfo {year} {2023})}\BibitemShut {NoStop}%
\bibitem [{\citenamefont {Kusumaatmaja}\ and\ \citenamefont {Lipowsky}(2011)}]{Kusumaatmaja2011}%
  \BibitemOpen
  \bibfield  {author} {\bibinfo {author} {\bibfnamefont {H.}~\bibnamefont {Kusumaatmaja}}\ and\ \bibinfo {author} {\bibfnamefont {R.}~\bibnamefont {Lipowsky}},\ }\bibfield  {title} {\bibinfo {title} {Droplet-induced budding transitions of membranes},\ }\href@noop {} {\bibfield  {journal} {\bibinfo  {journal} {Soft Matter}\ }\textbf {\bibinfo {volume} {7}},\ \bibinfo {pages} {6914} (\bibinfo {year} {2011})}\BibitemShut {NoStop}%
\bibitem [{\citenamefont {Anselmo}\ \emph {et~al.}(2015)\citenamefont {Anselmo}, \citenamefont {Zhang}, \citenamefont {Kumar}, \citenamefont {Vogus}, \citenamefont {Menegatti}, \citenamefont {Helgeson},\ and\ \citenamefont {Mitragotri}}]{anselmo2015elasticity}%
  \BibitemOpen
  \bibfield  {author} {\bibinfo {author} {\bibfnamefont {A.~C.}\ \bibnamefont {Anselmo}}, \bibinfo {author} {\bibfnamefont {M.}~\bibnamefont {Zhang}}, \bibinfo {author} {\bibfnamefont {S.}~\bibnamefont {Kumar}}, \bibinfo {author} {\bibfnamefont {D.~R.}\ \bibnamefont {Vogus}}, \bibinfo {author} {\bibfnamefont {S.}~\bibnamefont {Menegatti}}, \bibinfo {author} {\bibfnamefont {M.~E.}\ \bibnamefont {Helgeson}},\ and\ \bibinfo {author} {\bibfnamefont {S.}~\bibnamefont {Mitragotri}},\ }\bibfield  {title} {\bibinfo {title} {Elasticity of nanoparticles influences their blood circulation, phagocytosis, endocytosis, and targeting},\ }\href@noop {} {\bibfield  {journal} {\bibinfo  {journal} {ACS Nano}\ }\textbf {\bibinfo {volume} {9}},\ \bibinfo {pages} {3169} (\bibinfo {year} {2015})}\BibitemShut {NoStop}%
\bibitem [{\citenamefont {Sun}\ \emph {et~al.}(2014)\citenamefont {Sun}, \citenamefont {Zhang}, \citenamefont {Wang}, \citenamefont {Feng}, \citenamefont {Liu}, \citenamefont {Yin}, \citenamefont {Xu}, \citenamefont {Wei}, \citenamefont {Ding}, \citenamefont {Shi} \emph {et~al.}}]{sun2014tunable}%
  \BibitemOpen
  \bibfield  {author} {\bibinfo {author} {\bibfnamefont {J.}~\bibnamefont {Sun}}, \bibinfo {author} {\bibfnamefont {L.}~\bibnamefont {Zhang}}, \bibinfo {author} {\bibfnamefont {J.}~\bibnamefont {Wang}}, \bibinfo {author} {\bibfnamefont {Q.}~\bibnamefont {Feng}}, \bibinfo {author} {\bibfnamefont {D.}~\bibnamefont {Liu}}, \bibinfo {author} {\bibfnamefont {Q.}~\bibnamefont {Yin}}, \bibinfo {author} {\bibfnamefont {D.}~\bibnamefont {Xu}}, \bibinfo {author} {\bibfnamefont {Y.}~\bibnamefont {Wei}}, \bibinfo {author} {\bibfnamefont {B.}~\bibnamefont {Ding}}, \bibinfo {author} {\bibfnamefont {X.}~\bibnamefont {Shi}}, \emph {et~al.},\ }\bibfield  {title} {\bibinfo {title} {Tunable rigidity of (polymeric core)-(lipid shell) nanoparticles for regulated cellular uptake.},\ }\href@noop {} {\bibfield  {journal} {\bibinfo  {journal} {Advanced Materials}\ }\textbf {\bibinfo {volume} {27}},\ \bibinfo {pages} {1402} (\bibinfo {year} {2014})}\BibitemShut {NoStop}%
\bibitem [{\citenamefont {Dimova}\ and\ \citenamefont {Marques}(2019)}]{dimova2019giant}%
  \BibitemOpen
  \bibfield  {author} {\bibinfo {author} {\bibfnamefont {R.}~\bibnamefont {Dimova}}\ and\ \bibinfo {author} {\bibfnamefont {C.}~\bibnamefont {Marques}},\ }\href@noop {} {\emph {\bibinfo {title} {The giant vesicle book}}}\ (\bibinfo  {publisher} {CRC Press},\ \bibinfo {year} {2019})\BibitemShut {NoStop}%
\bibitem [{\citenamefont {Dinsmore}\ \emph {et~al.}(1998)\citenamefont {Dinsmore}, \citenamefont {Wong}, \citenamefont {Nelson},\ and\ \citenamefont {Yodh}}]{dinsmore1998hard}%
  \BibitemOpen
  \bibfield  {author} {\bibinfo {author} {\bibfnamefont {A.}~\bibnamefont {Dinsmore}}, \bibinfo {author} {\bibfnamefont {D.}~\bibnamefont {Wong}}, \bibinfo {author} {\bibfnamefont {P.}~\bibnamefont {Nelson}},\ and\ \bibinfo {author} {\bibfnamefont {A.}~\bibnamefont {Yodh}},\ }\bibfield  {title} {\bibinfo {title} {Hard spheres in vesicles: curvature-induced forces and particle-induced curvature},\ }\href@noop {} {\bibfield  {journal} {\bibinfo  {journal} {Physical Review Letters}\ }\textbf {\bibinfo {volume} {80}},\ \bibinfo {pages} {409} (\bibinfo {year} {1998})}\BibitemShut {NoStop}%
\bibitem [{\citenamefont {Asakura}\ and\ \citenamefont {Oosawa}(1954)}]{asakura1954interaction}%
  \BibitemOpen
  \bibfield  {author} {\bibinfo {author} {\bibfnamefont {S.}~\bibnamefont {Asakura}}\ and\ \bibinfo {author} {\bibfnamefont {F.}~\bibnamefont {Oosawa}},\ }\bibfield  {title} {\bibinfo {title} {On interaction between two bodies immersed in a solution of macromolecules},\ }\href@noop {} {\bibfield  {journal} {\bibinfo  {journal} {The Journal of Chemical Physics}\ }\textbf {\bibinfo {volume} {22}},\ \bibinfo {pages} {1255} (\bibinfo {year} {1954})}\BibitemShut {NoStop}%
\bibitem [{\citenamefont {Machado}\ \emph {et~al.}(2019)\citenamefont {Machado}, \citenamefont {Mercier},\ and\ \citenamefont {Chiaruttini}}]{machado2019limeseg}%
  \BibitemOpen
  \bibfield  {author} {\bibinfo {author} {\bibfnamefont {S.}~\bibnamefont {Machado}}, \bibinfo {author} {\bibfnamefont {V.}~\bibnamefont {Mercier}},\ and\ \bibinfo {author} {\bibfnamefont {N.}~\bibnamefont {Chiaruttini}},\ }\bibfield  {title} {\bibinfo {title} {Limeseg: a coarse-grained lipid membrane simulation for 3d image segmentation},\ }\href@noop {} {\bibfield  {journal} {\bibinfo  {journal} {BMC Bioinformatics}\ }\textbf {\bibinfo {volume} {20}},\ \bibinfo {pages} {1} (\bibinfo {year} {2019})}\BibitemShut {NoStop}%
\bibitem [{\citenamefont {Schindelin}\ \emph {et~al.}(2012)\citenamefont {Schindelin}, \citenamefont {Arganda-Carreras}, \citenamefont {Frise}, \citenamefont {Kaynig}, \citenamefont {Longair}, \citenamefont {Pietzsch}, \citenamefont {Preibisch}, \citenamefont {Rueden}, \citenamefont {Saalfeld}, \citenamefont {Schmid} \emph {et~al.}}]{schindelin2012fiji}%
  \BibitemOpen
  \bibfield  {author} {\bibinfo {author} {\bibfnamefont {J.}~\bibnamefont {Schindelin}}, \bibinfo {author} {\bibfnamefont {I.}~\bibnamefont {Arganda-Carreras}}, \bibinfo {author} {\bibfnamefont {E.}~\bibnamefont {Frise}}, \bibinfo {author} {\bibfnamefont {V.}~\bibnamefont {Kaynig}}, \bibinfo {author} {\bibfnamefont {M.}~\bibnamefont {Longair}}, \bibinfo {author} {\bibfnamefont {T.}~\bibnamefont {Pietzsch}}, \bibinfo {author} {\bibfnamefont {S.}~\bibnamefont {Preibisch}}, \bibinfo {author} {\bibfnamefont {C.}~\bibnamefont {Rueden}}, \bibinfo {author} {\bibfnamefont {S.}~\bibnamefont {Saalfeld}}, \bibinfo {author} {\bibfnamefont {B.}~\bibnamefont {Schmid}}, \emph {et~al.},\ }\bibfield  {title} {\bibinfo {title} {Fiji: an open-source platform for biological-image analysis},\ }\href@noop {} {\bibfield  {journal} {\bibinfo  {journal} {Nature Methods}\ }\textbf {\bibinfo {volume} {9}},\ \bibinfo {pages} {676} (\bibinfo {year} {2012})}\BibitemShut {NoStop}%
\bibitem [{\citenamefont {Seifert}\ \emph {et~al.}(1991)\citenamefont {Seifert}, \citenamefont {Berndl},\ and\ \citenamefont {Lipowsky}}]{seifert1991shape}%
  \BibitemOpen
  \bibfield  {author} {\bibinfo {author} {\bibfnamefont {U.}~\bibnamefont {Seifert}}, \bibinfo {author} {\bibfnamefont {K.}~\bibnamefont {Berndl}},\ and\ \bibinfo {author} {\bibfnamefont {R.}~\bibnamefont {Lipowsky}},\ }\bibfield  {title} {\bibinfo {title} {Shape transformations of vesicles: Phase diagram for spontaneous-curvature and bilayer-coupling models},\ }\href@noop {} {\bibfield  {journal} {\bibinfo  {journal} {Physical Review A}\ }\textbf {\bibinfo {volume} {44}},\ \bibinfo {pages} {1182} (\bibinfo {year} {1991})}\BibitemShut {NoStop}%
\bibitem [{\citenamefont {Faizi}\ \emph {et~al.}(2020)\citenamefont {Faizi}, \citenamefont {Reeves}, \citenamefont {Georgiev}, \citenamefont {Vlahovska},\ and\ \citenamefont {Dimova}}]{faizi2020fluctuation}%
  \BibitemOpen
  \bibfield  {author} {\bibinfo {author} {\bibfnamefont {H.~A.}\ \bibnamefont {Faizi}}, \bibinfo {author} {\bibfnamefont {C.~J.}\ \bibnamefont {Reeves}}, \bibinfo {author} {\bibfnamefont {V.~N.}\ \bibnamefont {Georgiev}}, \bibinfo {author} {\bibfnamefont {P.~M.}\ \bibnamefont {Vlahovska}},\ and\ \bibinfo {author} {\bibfnamefont {R.}~\bibnamefont {Dimova}},\ }\bibfield  {title} {\bibinfo {title} {Fluctuation spectroscopy of giant unilamellar vesicles using confocal and phase contrast microscopy},\ }\href@noop {} {\bibfield  {journal} {\bibinfo  {journal} {Soft Matter}\ }\textbf {\bibinfo {volume} {16}},\ \bibinfo {pages} {8996} (\bibinfo {year} {2020})}\BibitemShut {NoStop}%
\bibitem [{\citenamefont {Francois}\ \emph {et~al.}(1979)\citenamefont {Francois}, \citenamefont {Sarazin}, \citenamefont {Schwartz},\ and\ \citenamefont {Weill}}]{francois1979polyacrylamide}%
  \BibitemOpen
  \bibfield  {author} {\bibinfo {author} {\bibfnamefont {J.}~\bibnamefont {Francois}}, \bibinfo {author} {\bibfnamefont {D.}~\bibnamefont {Sarazin}}, \bibinfo {author} {\bibfnamefont {T.}~\bibnamefont {Schwartz}},\ and\ \bibinfo {author} {\bibfnamefont {G.}~\bibnamefont {Weill}},\ }\bibfield  {title} {\bibinfo {title} {Polyacrylamide in water: molecular weight dependence of $<\mathrm{R}2>$ and [$\eta$] and the problem of the excluded volume exponent},\ }\href@noop {} {\bibfield  {journal} {\bibinfo  {journal} {Polymer}\ }\textbf {\bibinfo {volume} {20}},\ \bibinfo {pages} {969} (\bibinfo {year} {1979})}\BibitemShut {NoStop}%
\bibitem [{\citenamefont {Tuinier}\ and\ \citenamefont {Lekkerkerker}(2001)}]{tuinier2001excluded}%
  \BibitemOpen
  \bibfield  {author} {\bibinfo {author} {\bibfnamefont {R.}~\bibnamefont {Tuinier}}\ and\ \bibinfo {author} {\bibfnamefont {H.}~\bibnamefont {Lekkerkerker}},\ }\bibfield  {title} {\bibinfo {title} {Excluded-volume polymer-induced depletion interaction between parallel flat plates},\ }\href@noop {} {\bibfield  {journal} {\bibinfo  {journal} {The European Physical Journal E}\ }\textbf {\bibinfo {volume} {6}},\ \bibinfo {pages} {129} (\bibinfo {year} {2001})}\BibitemShut {NoStop}%
\bibitem [{\citenamefont {Helfrich}\ and\ \citenamefont {Servuss}(1984)}]{helfrich1984undulations}%
  \BibitemOpen
  \bibfield  {author} {\bibinfo {author} {\bibfnamefont {W.}~\bibnamefont {Helfrich}}\ and\ \bibinfo {author} {\bibfnamefont {R.~M.}\ \bibnamefont {Servuss}},\ }\bibfield  {title} {\bibinfo {title} {Undulations, steric interaction and cohesion of fluid membranes},\ }\href@noop {} {\bibfield  {journal} {\bibinfo  {journal} {Il Nuovo Cimento D}\ }\textbf {\bibinfo {volume} {3}},\ \bibinfo {pages} {137} (\bibinfo {year} {1984})}\BibitemShut {NoStop}%
\bibitem [{\citenamefont {Lipowsky}\ and\ \citenamefont {Sackmann}(1995)}]{lipowsky1995structure}%
  \BibitemOpen
  \bibfield  {author} {\bibinfo {author} {\bibfnamefont {R.}~\bibnamefont {Lipowsky}}\ and\ \bibinfo {author} {\bibfnamefont {E.}~\bibnamefont {Sackmann}},\ }\href@noop {} {\emph {\bibinfo {title} {Structure and dynamics of membranes: I. from cells to vesicles/II. generic and specific interactions}}}\ (\bibinfo  {publisher} {Elsevier},\ \bibinfo {year} {1995})\BibitemShut {NoStop}%
\bibitem [{\citenamefont {Deng}\ \emph {et~al.}(2017)\citenamefont {Deng}, \citenamefont {Yelleswarapu}, \citenamefont {Zheng},\ and\ \citenamefont {Huck}}]{deng2017microfluidic}%
  \BibitemOpen
  \bibfield  {author} {\bibinfo {author} {\bibfnamefont {N.-N.}\ \bibnamefont {Deng}}, \bibinfo {author} {\bibfnamefont {M.}~\bibnamefont {Yelleswarapu}}, \bibinfo {author} {\bibfnamefont {L.}~\bibnamefont {Zheng}},\ and\ \bibinfo {author} {\bibfnamefont {W.~T.}\ \bibnamefont {Huck}},\ }\bibfield  {title} {\bibinfo {title} {Microfluidic assembly of monodisperse vesosomes as artificial cell models},\ }\href@noop {} {\bibfield  {journal} {\bibinfo  {journal} {Journal of the American Chemical Society}\ }\textbf {\bibinfo {volume} {139}},\ \bibinfo {pages} {587} (\bibinfo {year} {2017})}\BibitemShut {NoStop}%
\bibitem [{\citenamefont {Kraus}\ \emph {et~al.}(1995)\citenamefont {Kraus}, \citenamefont {Seifert},\ and\ \citenamefont {Lipowsky}}]{kraus1995gravity}%
  \BibitemOpen
  \bibfield  {author} {\bibinfo {author} {\bibfnamefont {M.}~\bibnamefont {Kraus}}, \bibinfo {author} {\bibfnamefont {U.}~\bibnamefont {Seifert}},\ and\ \bibinfo {author} {\bibfnamefont {R.}~\bibnamefont {Lipowsky}},\ }\bibfield  {title} {\bibinfo {title} {Gravity-induced shape transformations of vesicles},\ }\href@noop {} {\bibfield  {journal} {\bibinfo  {journal} {Europhysics Letters}\ }\textbf {\bibinfo {volume} {32}},\ \bibinfo {pages} {431} (\bibinfo {year} {1995})}\BibitemShut {NoStop}%
\bibitem [{\citenamefont {Pernpeintner}\ \emph {et~al.}(2017)\citenamefont {Pernpeintner}, \citenamefont {Frank}, \citenamefont {Urban}, \citenamefont {Roeske}, \citenamefont {Pritzl}, \citenamefont {Trauner},\ and\ \citenamefont {Lohmu{\"u}ller}}]{pernpeintner2017light}%
  \BibitemOpen
  \bibfield  {author} {\bibinfo {author} {\bibfnamefont {C.}~\bibnamefont {Pernpeintner}}, \bibinfo {author} {\bibfnamefont {J.~A.}\ \bibnamefont {Frank}}, \bibinfo {author} {\bibfnamefont {P.}~\bibnamefont {Urban}}, \bibinfo {author} {\bibfnamefont {C.~R.}\ \bibnamefont {Roeske}}, \bibinfo {author} {\bibfnamefont {S.~D.}\ \bibnamefont {Pritzl}}, \bibinfo {author} {\bibfnamefont {D.}~\bibnamefont {Trauner}},\ and\ \bibinfo {author} {\bibfnamefont {T.}~\bibnamefont {Lohmu{\"u}ller}},\ }\bibfield  {title} {\bibinfo {title} {Light-controlled membrane mechanics and shape transitions of photoswitchable lipid vesicles},\ }\href@noop {} {\bibfield  {journal} {\bibinfo  {journal} {Langmuir}\ }\textbf {\bibinfo {volume} {33}},\ \bibinfo {pages} {4083} (\bibinfo {year} {2017})}\BibitemShut {NoStop}%
\bibitem [{\citenamefont {Aleksanyan}\ \emph {et~al.}(2023)\citenamefont {Aleksanyan}, \citenamefont {Grafm{\"u}ller}, \citenamefont {Crea}, \citenamefont {Georgiev}, \citenamefont {Yandrapalli}, \citenamefont {Block}, \citenamefont {Heberle},\ and\ \citenamefont {Dimova}}]{aleksanyan2023photomanipulation}%
  \BibitemOpen
  \bibfield  {author} {\bibinfo {author} {\bibfnamefont {M.}~\bibnamefont {Aleksanyan}}, \bibinfo {author} {\bibfnamefont {A.}~\bibnamefont {Grafm{\"u}ller}}, \bibinfo {author} {\bibfnamefont {F.}~\bibnamefont {Crea}}, \bibinfo {author} {\bibfnamefont {V.~N.}\ \bibnamefont {Georgiev}}, \bibinfo {author} {\bibfnamefont {N.}~\bibnamefont {Yandrapalli}}, \bibinfo {author} {\bibfnamefont {S.}~\bibnamefont {Block}}, \bibinfo {author} {\bibfnamefont {J.}~\bibnamefont {Heberle}},\ and\ \bibinfo {author} {\bibfnamefont {R.}~\bibnamefont {Dimova}},\ }\bibfield  {title} {\bibinfo {title} {Photomanipulation of minimal synthetic cells: Area increase, softening, and interleaflet coupling of membrane models doped with azobenzene-lipid photoswitches},\ }\href@noop {} {\bibfield  {journal} {\bibinfo  {journal} {Advanced Science}\ }\textbf {\bibinfo {volume} {10}},\ \bibinfo {pages} {2304336} (\bibinfo {year} {2023})}\BibitemShut {NoStop}%
\bibitem [{\citenamefont {Mangiarotti}\ \emph {et~al.}(2024)\citenamefont {Mangiarotti}, \citenamefont {Aleksanyan}, \citenamefont {Siri}, \citenamefont {Sun}, \citenamefont {Lipowsky},\ and\ \citenamefont {Dimova}}]{mangiarotti2024photoswitchable}%
  \BibitemOpen
  \bibfield  {author} {\bibinfo {author} {\bibfnamefont {A.}~\bibnamefont {Mangiarotti}}, \bibinfo {author} {\bibfnamefont {M.}~\bibnamefont {Aleksanyan}}, \bibinfo {author} {\bibfnamefont {M.}~\bibnamefont {Siri}}, \bibinfo {author} {\bibfnamefont {T.-W.}\ \bibnamefont {Sun}}, \bibinfo {author} {\bibfnamefont {R.}~\bibnamefont {Lipowsky}},\ and\ \bibinfo {author} {\bibfnamefont {R.}~\bibnamefont {Dimova}},\ }\bibfield  {title} {\bibinfo {title} {Photoswitchable endocytosis of biomolecular condensates in giant vesicles},\ }\href@noop {} {\bibfield  {journal} {\bibinfo  {journal} {Advanced Science}\ }\textbf {\bibinfo {volume} {11}},\ \bibinfo {pages} {2309864} (\bibinfo {year} {2024})}\BibitemShut {NoStop}%
\bibitem [{\citenamefont {Agudo-Canalejo}(2020)}]{agudo2020engulfment}%
  \BibitemOpen
  \bibfield  {author} {\bibinfo {author} {\bibfnamefont {J.}~\bibnamefont {Agudo-Canalejo}},\ }\bibfield  {title} {\bibinfo {title} {Engulfment of ellipsoidal nanoparticles by membranes: full description of orientational changes},\ }\href@noop {} {\bibfield  {journal} {\bibinfo  {journal} {Journal of Physics: Condensed Matter}\ }\textbf {\bibinfo {volume} {32}},\ \bibinfo {pages} {294001} (\bibinfo {year} {2020})}\BibitemShut {NoStop}%
\bibitem [{\citenamefont {Vutukuri}\ \emph {et~al.}(2020)\citenamefont {Vutukuri}, \citenamefont {Hoore}, \citenamefont {Abaurrea-Velasco}, \citenamefont {van Buren}, \citenamefont {Dutto}, \citenamefont {Auth}, \citenamefont {Fedosov}, \citenamefont {Gompper},\ and\ \citenamefont {Vermant}}]{vutukuri2020active}%
  \BibitemOpen
  \bibfield  {author} {\bibinfo {author} {\bibfnamefont {H.~R.}\ \bibnamefont {Vutukuri}}, \bibinfo {author} {\bibfnamefont {M.}~\bibnamefont {Hoore}}, \bibinfo {author} {\bibfnamefont {C.}~\bibnamefont {Abaurrea-Velasco}}, \bibinfo {author} {\bibfnamefont {L.}~\bibnamefont {van Buren}}, \bibinfo {author} {\bibfnamefont {A.}~\bibnamefont {Dutto}}, \bibinfo {author} {\bibfnamefont {T.}~\bibnamefont {Auth}}, \bibinfo {author} {\bibfnamefont {D.~A.}\ \bibnamefont {Fedosov}}, \bibinfo {author} {\bibfnamefont {G.}~\bibnamefont {Gompper}},\ and\ \bibinfo {author} {\bibfnamefont {J.}~\bibnamefont {Vermant}},\ }\bibfield  {title} {\bibinfo {title} {Active particles induce large shape deformations in giant lipid vesicles},\ }\href@noop {} {\bibfield  {journal} {\bibinfo  {journal} {Nature}\ }\textbf {\bibinfo {volume} {586}},\ \bibinfo {pages} {52} (\bibinfo {year} {2020})}\BibitemShut {NoStop}%
\bibitem [{\citenamefont {Moga}\ \emph {et~al.}(2019)\citenamefont {Moga}, \citenamefont {Yandrapalli}, \citenamefont {Dimova},\ and\ \citenamefont {Robinson}}]{moga2019optimization}%
  \BibitemOpen
  \bibfield  {author} {\bibinfo {author} {\bibfnamefont {A.}~\bibnamefont {Moga}}, \bibinfo {author} {\bibfnamefont {N.}~\bibnamefont {Yandrapalli}}, \bibinfo {author} {\bibfnamefont {R.}~\bibnamefont {Dimova}},\ and\ \bibinfo {author} {\bibfnamefont {T.}~\bibnamefont {Robinson}},\ }\bibfield  {title} {\bibinfo {title} {Optimization of the inverted emulsion method for high-yield production of biomimetic giant unilamellar vesicles},\ }\href@noop {} {\bibfield  {journal} {\bibinfo  {journal} {ChemBioChem}\ }\textbf {\bibinfo {volume} {20}},\ \bibinfo {pages} {2674} (\bibinfo {year} {2019})}\BibitemShut {NoStop}%
\bibitem [{\citenamefont {Frank}\ \emph {et~al.}(2016)\citenamefont {Frank}, \citenamefont {Yushchenko}, \citenamefont {Hodson}, \citenamefont {Lipstein}, \citenamefont {Nagpal}, \citenamefont {Rutter}, \citenamefont {Rhee}, \citenamefont {Gottschalk}, \citenamefont {Brose}, \citenamefont {Schultz} \emph {et~al.}}]{frank2016photoswitchable}%
  \BibitemOpen
  \bibfield  {author} {\bibinfo {author} {\bibfnamefont {J.~A.}\ \bibnamefont {Frank}}, \bibinfo {author} {\bibfnamefont {D.~A.}\ \bibnamefont {Yushchenko}}, \bibinfo {author} {\bibfnamefont {D.~J.}\ \bibnamefont {Hodson}}, \bibinfo {author} {\bibfnamefont {N.}~\bibnamefont {Lipstein}}, \bibinfo {author} {\bibfnamefont {J.}~\bibnamefont {Nagpal}}, \bibinfo {author} {\bibfnamefont {G.~A.}\ \bibnamefont {Rutter}}, \bibinfo {author} {\bibfnamefont {J.-S.}\ \bibnamefont {Rhee}}, \bibinfo {author} {\bibfnamefont {A.}~\bibnamefont {Gottschalk}}, \bibinfo {author} {\bibfnamefont {N.}~\bibnamefont {Brose}}, \bibinfo {author} {\bibfnamefont {C.}~\bibnamefont {Schultz}}, \emph {et~al.},\ }\bibfield  {title} {\bibinfo {title} {Photoswitchable diacylglycerols enable optical control of protein kinase c},\ }\href@noop {} {\bibfield  {journal} {\bibinfo  {journal} {Nature Chemical Biology}\ }\textbf {\bibinfo {volume} {12}},\ \bibinfo {pages} {755} (\bibinfo {year} {2016})}\BibitemShut {NoStop}%
\bibitem [{\citenamefont {Weakly}\ \emph {et~al.}(2024)\citenamefont {Weakly}, \citenamefont {Wilson}, \citenamefont {Goetz}, \citenamefont {Pruitt}, \citenamefont {Li}, \citenamefont {Xu},\ and\ \citenamefont {Keller}}]{weakly2024several}%
  \BibitemOpen
  \bibfield  {author} {\bibinfo {author} {\bibfnamefont {H.~M.}\ \bibnamefont {Weakly}}, \bibinfo {author} {\bibfnamefont {K.~J.}\ \bibnamefont {Wilson}}, \bibinfo {author} {\bibfnamefont {G.~J.}\ \bibnamefont {Goetz}}, \bibinfo {author} {\bibfnamefont {E.~L.}\ \bibnamefont {Pruitt}}, \bibinfo {author} {\bibfnamefont {A.}~\bibnamefont {Li}}, \bibinfo {author} {\bibfnamefont {L.}~\bibnamefont {Xu}},\ and\ \bibinfo {author} {\bibfnamefont {S.~L.}\ \bibnamefont {Keller}},\ }\bibfield  {title} {\bibinfo {title} {Several common methods of making vesicles (except an emulsion method) capture intended lipid ratios},\ }\href@noop {} {\bibfield  {journal} {\bibinfo  {journal} {Biophysical Journal}\ }\textbf {\bibinfo {volume} {123}},\ \bibinfo {pages} {3452} (\bibinfo {year} {2024})}\BibitemShut {NoStop}%
\bibitem [{\citenamefont {Diel}\ \emph {et~al.}(2020)\citenamefont {Diel}, \citenamefont {Lichtman},\ and\ \citenamefont {Richardson}}]{diel2020tutorial}%
  \BibitemOpen
  \bibfield  {author} {\bibinfo {author} {\bibfnamefont {E.~E.}\ \bibnamefont {Diel}}, \bibinfo {author} {\bibfnamefont {J.~W.}\ \bibnamefont {Lichtman}},\ and\ \bibinfo {author} {\bibfnamefont {D.~S.}\ \bibnamefont {Richardson}},\ }\bibfield  {title} {\bibinfo {title} {Tutorial: avoiding and correcting sample-induced spherical aberration artifacts in 3d fluorescence microscopy},\ }\href@noop {} {\bibfield  {journal} {\bibinfo  {journal} {Nature Protocols}\ }\textbf {\bibinfo {volume} {15}},\ \bibinfo {pages} {2773} (\bibinfo {year} {2020})}\BibitemShut {NoStop}%
\bibitem [{\citenamefont {Jacobson}\ \emph {et~al.}(2018)\citenamefont {Jacobson}, \citenamefont {Panozzo} \emph {et~al.}}]{libigl}%
  \BibitemOpen
  \bibfield  {author} {\bibinfo {author} {\bibfnamefont {A.}~\bibnamefont {Jacobson}}, \bibinfo {author} {\bibfnamefont {D.}~\bibnamefont {Panozzo}}, \emph {et~al.},\ }\href@noop {} {\bibinfo {title} {{libigl}: A simple {C++} geometry processing library}} (\bibinfo {year} {2018}),\ \bibinfo {note} {https://libigl.github.io/}\BibitemShut {NoStop}%
\end{thebibliography}%

\end{document}